\documentstyle[12pt,subeqn,psfig,a4]{article}

\advance\textwidth by 100pt
\advance\oddsidemargin by -50pt
\advance\evensidemargin by -50pt

\def \ie {{\it i.e.}}
\renewcommand{\thesection}{\arabic{section}}
\renewcommand{\theequation}{\thesection.\arabic{equation}}
\newcommand{\cleqn}{\setcounter{equation}{0}}

\newcommand {\hate}     {\hat{e}}

\newcommand {\hats}     {\hat{s}}
\newcommand {\hatc}     {\hat{c}}
\newcommand {\hatgz}    {\hat{g}_Z}

\newcommand {\hatesq}   {\hat{e}^2}
\newcommand {\hatgsq}   {\hat{g}^2}
\newcommand {\hatssq}   {\hat{s}^2}
\newcommand {\hatcsq}   {\hat{c}^2}
\newcommand {\hatgzsq}  {\hat{g}_Z^2}

\newcommand {\mwsq}  {m_W^2}
\newcommand {\mzsq}  {m_Z^2}

\newcommand{\scfer}{{\tilde{f}}}

\newcommand{\scferl}{{\tilde{f}_L}}
\newcommand{\scferr}{{\tilde{f}_R}}

\newcommand{\scup}{{\tilde{u}}}

\newcommand{\scupl}{{\tilde{u}_L}}
\newcommand{\scupr}{{\tilde{u}_R}}

\newcommand{\scdown}{{\tilde{d}}}

\newcommand{\scdownl}{{\tilde{d}_L}}
\newcommand{\scdownr}{{\tilde{d}_R}}

\newcommand {\pigg}    {\Pi^{\gamma\gamma}}
\newcommand {\pigz}    {\Pi^{\gamma Z}}
\newcommand {\pizz}    {\Pi^{ZZ}}
\newcommand {\piww}    {\Pi^{WW}}

\newcommand {\pitgg}    {\Pi_T^{\gamma\gamma}}
\newcommand {\pitgz}    {\Pi_T^{\gamma Z}}
\newcommand {\pitzz}    {\Pi_T^{ZZ}}
\newcommand {\pitww}    {\Pi_T^{WW}}

\newcommand {\pitggg} {{\Pi}_{T,\gamma}^{\gamma\gamma}}

\newcommand {\pitgzg} {{\Pi}_{T,\gamma}^{\gamma Z}}
\newcommand {\pitzzz} {{\Pi}_{T,Z}^{ZZ}}

\newcommand {\qsq} {{q^2}}

\newcommand {\HHbar}{{\stackrel{{\scriptscriptstyle (}-{\scriptscriptstyle )}}{H}}}
\newcommand {\hhbar}{{\stackrel{{\scriptscriptstyle (}-{\scriptscriptstyle )}}{h}}}
\newcommand {\GGbar}{{\stackrel{{\scriptscriptstyle (}-{\scriptscriptstyle )}}{G}}}
\newcommand {\xixibar}{{\stackrel{{\scriptscriptstyle (}-{\scriptscriptstyle )}}{\xi}}}
\newcommand {\born}{{\rm [Born]}}
\newcommand {\nc}  {{\rm [NC]}}
\newcommand {\gvtx}{{{\rm [}\gamma{\rm, VTX]}}}
\newcommand {\zvtx}{{{\rm [}Z{\rm, VTX]}}}

\newcommand {\boxes}  {{\rm [Box]}}
\newcommand {\wphys}   {{{\rm [}W_P{\rm]}}}
\newcommand {\unphys}   {{{\rm [}W_S,\chi{\rm]}}}

\begin{document}

\title{Sum rules for $e^+e^- \rightarrow W^+W^-$ helicity 
amplitudes\\ from BRS invariance}
\author{S.~Alam$^{1,2}$, K.~Hagiwara$^{1}$, S.~Kanemura$^1$,\\
          R.~Szalapski$^1$ and Y.~Umeda$^{1,3}$ }
\date{}
\maketitle
\begin{center}
$^1$Theory Group, KEK, Tsukuba, Ibaraki 305-0801, Japan\\
$^2$Physics Department, University of Peshawar, Peshawar, NWFP, Pakistan\\
$^3$Department of Physics, Hokaido University, Sapporo 060-0010, Japan  
\end{center}


\vspace*{-9cm}\hspace*{12cm}
\vbox{\baselineskip14pt
\hbox{\bf KEK Preprint 98-128}
\hbox{\bf KEK-TH-573}
\hbox{\bf EPHOU-98-010}
}
\vspace*{10cm}

\begin{abstract}
The BRS invariance of the electroweak gauge theory leads to relationships 
between amplitudes with external massive gauge bosons and amplitudes where 
some of these gauge bosons are replaced with their corresponding 
Nambu-Goldstone bosons.  Unlike the equivalence theorem, these identities 
are exact at all energies.  In this paper we discuss such identities which 
relate the process $e^+e^- \rightarrow W^+W^-$ to $W^\pm\chi^\mp$ and 
$\chi^+\chi^-$ production.  By using a general form-factor decomposition for
$e^+e^- \rightarrow W^+W^-$, $e^+e^- \rightarrow  W^\pm \chi^\mp$ and 
$e^+e^- \rightarrow \chi^+\chi^-$ amplitudes, these identities are expressed 
as sum rules among scalar form factors.   Because these sum rules may be 
applied order by order in perturbation theory, they provide a powerful test of 
higher order calculations.  By using additional Ward-Takahashi identities
we find that the various contributions are divided into separately 
gauge-invariant subsets, the sum rules applying independently to each subset.
After a general discussion of the application of the sum rules we consider the 
one-loop contributions of scalar-fermions in the Minimal Supersymmetric 
Standard Model as an illustration.
\end{abstract}

\newpage

\section{Introduction}
\label{sec-intro}
\cleqn

Particle physicists devote an enormous amount of time and energy to the 
calculation of loop effects in perturbation theory.  The calculations can be 
formidable.  Even when the analytical portion of the calculation is 
completely correct,
the numerical evaluation of those expressions on a computer can lead to 
additional difficulties.  
For example, the numerical evaluation of loop integrals can be problematic 
in various kinematical regions. 
Of particular interest in this paper are the complications that arise when 
gauge cancellations take place between the various Feynman diagrams that 
contribute to an amplitude. 
A careless treatment of higher order effects 
or round-off errors can lead to numerical violations of the gauge 
cancellation.
The magnitudes of these errors can grow with energy becoming very serious 
at higher energies.  
Hence, it is common for difficult calculations to 
be performed by two or more independent collaborations in the hope that 
the redundancy will allow errors to be eliminated.  
Any tools which allow the practitioner to check his or her own results 
independently of other calculations 
are thus of great value.  In this paper we study one such tool, sum rules 
among form factors that follow from BRS symmetry.

The standard electroweak theory after gauge fixing is invariant under a global 
BRS\cite{brs} symmetry.  As a result, amplitudes which include external 
massive gauge bosons may be related to amplitudes where some of those bosons 
are replaced with their Nambu-Goldstone counterparts.  The identities may be
derived formally by noting that the gauge fixing term is generated by the BRS 
transformation of the anti-ghost field $\overline{c}$,  
\begin{equation}
\label{anticommutator}
  \left\{ Q_{\rm BRS}, \widehat{\overline{c}}^\pm \right\} = 
  \partial^\mu \widehat{W}_\mu^\pm 
   + \hat{\xi}_W \hat{m}_W \widehat{\chi}^\pm\;,
\end{equation}
where $\widehat{W}^\pm$ and $\widehat{\chi}^\pm$ are the $\overline{\rm MS}$ 
operators for the weak gauge bosons and the corresponding Goldstone bosons,
respectively, and 
$\hat{\xi}_W$ and $\hat{m}_W$ are the $\overline{\rm MS}$ gauge-fixing 
parameter and the $\overline{\rm MS}$ $W$-boson mass, respectively. 
  Then, from the condition that physical states are annihilated by 
the BRS charge,
\begin{equation}
  Q_{\rm BRS} | {\rm phys} \rangle = 
  \langle {\rm phys} | Q_{\rm BRS} = 0\;,
\end{equation}
we obtain the following identity\cite{gkn86}:
\begin{equation}
 \langle {\rm out}\, | \, (\partial^\mu \widehat{W}^\pm_\mu 
 + \hat{\xi}_W \hat{m}_W \widehat{\chi}^\pm) \, 
   |\, {\rm in} \rangle = 0\;,
\label{brsid}  
\end{equation}
where $|\, {\rm in} \rangle$ and $|\, {\rm out} \rangle$ denote arbitrary 
incoming and outgoing physical states, respectively.  

The matrix elements which include 
an outgoing asymptotic state of an unphysical 
field may be expressed by 
\begin{eqnarray}
&&
\!\!\!\!\!\!\!\!\!\!\!\!\!\!\!\!\!\!\!\!\!\!
\langle {\rm out }| a_{W^\pm \rm out}(\vec{k},S) | {\rm in} \rangle  
= -Z_W^{1/2} Z_m 
\int \frac{d^4 x}{\sqrt{(2\pi)^3 2 k_0}} 
 \left( \frac{1}{\hat{\xi}_W \hat{m}_W} \right) \;e^{i k x}  
          \left(\Box + \xi_W m_W^2 \right)    
        \langle {\rm out }| 
   \partial^\mu \widehat{W}_\mu^\pm (x)  | {\rm in} \rangle \;, 
\label{mtrx1} \\   
&&\!\!\!\!\!\!\!\!\!\!\!\!\!\!\!
\langle {\rm out} | a_{\chi^\pm \;{\rm out}}(\vec{k}) | {\rm in} \rangle 
 = i Z_\chi^{-1/2} \int \frac{d^4 x}{\sqrt{(2\pi)^3 2 k_0}} \; e^{i k x} 
\left( \Box + \xi_W m_W^2 \right)
    \langle {\rm out} | \widehat{\chi}^\pm(x) | {\rm in} \rangle\;,
\label{mtrx2}  
\end{eqnarray}
where $a_{W^\pm \rm out}(\vec{k},S)$ and $a_{\chi^\pm \;{\rm out}}(\vec{k})$ 
are the annihilation operators for the renormalized scalar $W$ boson,  
$W^\pm_S$, and the renormalized Nambu-Goldstone boson, $\chi^\pm$, 
respectively. 
The renormalized fields are related to 
to the $\overline{\rm MS}$ fields by  
$\partial^\mu \widehat{W}_\mu^\pm = Z_W^{1/2} \partial^\mu W_\mu^\pm$ and 
$\widehat{\chi}^\pm = Z_\chi^{1/2} \chi^\pm$.
The physical mass of $W$ bosons, $m_W$, and the renormalized gauge-fixing 
parameter, $\xi_W$, are parametrized as 
$\hat{m}_W = Z_m m_W$ and $\hat{\xi}_W = Z_W \xi_W$.  
From Eqns.~(\ref{brsid}), (\ref{mtrx1}) and (\ref{mtrx2}), we obtain 
\begin{eqnarray}
    \langle {\rm out} | a_{W^\pm\;{\rm out}} (\vec{k}, S) | {\rm in} \rangle 
  + i \,C_{\rm mod} 
    \langle {\rm out} | a_{\chi^\pm\;\rm out}(\vec{k})| {\rm in} \rangle
    \, = \,0 \;, \label{brsmtrx}
\end{eqnarray}
where 
\begin{eqnarray}
C_{\rm mod} =  Z_\chi^{1/2} Z_W^{1/2}\, Z_m \;.     \label{cmod}
\end{eqnarray}
The factor $C_{\rm mod}$ is unity at the tree level but receives  
corrections at higher-orders of perturbation theory. 
Hence, matrix elements where one external $W$-boson has 
an unphysical scalar polarization are related to amplitudes where that same
$W$-boson is replaced by its corresponding Goldstone boson.

One process of considerable physical interest is $W$-boson pair production
from fermion-pair annihilation.
In the following sections we study relations between the helicity 
amplitudes for $e^+e^- \rightarrow W^+W^-$ and those 
for $e^+e^- \rightarrow W^\pm \chi^\mp$:     
\begin{subequations}
\begin{eqnarray}
\label{eq-brs1}
{\cal M}\Big(e^+ e^- \rightarrow W^-_P\, W^+_S\Big) + i C_{\rm mod}
{\cal M}\Big(e^+ e^- \rightarrow W^-_P\, \chi^+\Big) &=& 0\;, \\
\label{eq-brs2}
{\cal M}\Big(e^+ e^- \rightarrow W^-_S\, W^+_P\Big) + i C_{\rm mod}
{\cal M}\Big(e^+ e^- \rightarrow \chi^-\, W^+_P\Big) &=& 0 \;.
\end{eqnarray}
\end{subequations}
Here $P$ denotes a physical $W$-boson helicity while $S$ denotes a 
scalar polarization. 
We will refer to these as `single' BRS identities since 
they are obtained from the identity (\ref{brsmtrx}) {\em via} 
a single insertion of the anti-commutator in 
Eqn.~(\ref{anticommutator}).  
We will also consider the `double' BRS identity,
\begin{eqnarray}
\nonumber
\lefteqn{{\cal M} (e^+e^- \rightarrow W^+_S\,W^-_S) 
 + i C_{\rm mod} {\cal M} (e^+e^- \rightarrow W^+_S\, \chi^-)}&&\\&&\mbox{}
 + i C_{\rm mod} {\cal M} (e^+e^- \rightarrow \chi^+ \,W^-_S)
 - \left\{ C_{\rm mod} \right\}^2 
{\cal M} (e^+e^- \rightarrow \chi^+ \,\chi^-) = 0 \;,
\label{doublebrs}
\end{eqnarray}
which is obtained by a double insertion of the anti-commutator.

By using a general form-factor decomposition of the $e^+ e^- \rightarrow 
W^+W^-$\cite{hhis97},  $e^+ e^- \rightarrow W^\mp\chi^\pm$ and 
$e^+ e^- \rightarrow \chi^+\chi^-$ amplitudes the BRS identities of 
Eqns.~(\ref{eq-brs1}), (\ref{eq-brs2}) and (\ref{doublebrs}) are expressed
as sum rules among the form factors for the different processes.  Since the 
BRS invariance of the scattering amplitudes is a nonperturbative property, 
the BRS sum rules hold at each order of the perturbative expansion.  
The BRS identity in the form of Eqn.~(\ref{brsmtrx}) 
has already been employed at the tree level, 
for example, by the authors and users of HELAS\cite{helas} and 
MadGraph\cite{madgraph} for the evaluation of 
complicated tree-level amplitudes\cite{himmz91}. We will extend this approach 
to one-loop calculations.  Note that, while the equivalence 
theorem\cite{et,et2,et3} is an approximation which is valid only in the high 
energy limit, our sum rules are exact relations at all energies. 

This paper is organized as follows.  The form-factor decompositions of 
$e^+ e^- \rightarrow W^+W^-$,  $e^+ e^- \rightarrow W^\mp\chi^\pm$ and 
$e^+ e^- \rightarrow \chi^+\chi^-$ amplitudes are given in 
Section~\ref{subsec-amp-eeww}, Section~\ref{subsec-amp-eewx} and 
Section~\ref{subsec-amp-eexx}, respectively.  
We then derive sum rules among the various form factors in 
Section~\ref{sec-brs}. Sections~\ref{subsec-ff-eeww} 
and \ref{subsec-ff-eewx} present notation for the perturbative calculation of 
the form factors for $W$-boson pair production and $W$-boson Goldstone-boson 
associated production; these expressions will be utilized throughout the paper.
  Section~\ref{sec-tree} is devoted to a discussion of
BRS invariance at the tree level.  The discussion is extended to the one-loop
level in Section~\ref{sec-brs-loop}.  While we begin with a discussion of 
relationships between full amplitudes, the many contributions are split into
invariant subsets; each of these subsets corresponds to an additional Ward 
identity.  We illustrate these points in Section~\ref{sec-brs-sfermion}
with a calculation of the one-loop contributions of an SU(2) doublet of scalar 
fermions (the squarks or sleptons of the Minimal Supersymmetric Standard 
Model).  In the final section we present our conclusions.

For clarity many of the detailed calculations have been relegated to the 
appendices.  Scalar-fermion contributions to the gauge-boson propagators
are given in Appendix~\ref{app-propagators}.  The $W_S^\pm$ and $\chi^\pm$
propagator corrections form a gauge-invariant subset and decouple from the 
main discussion; this is shown in Appendix~\ref{app-unphysical}.
In Appendix~\ref{app-vertices} we present explicit calculations of the form 
factors at one loop, while many of the detailed calculations concerning the 
BRS sum rules at one loop are given in Appendix~\ref{app-details}.  Finally,
we present our decomposition of the rank-three three-point tensor integral
to scalar integrals\cite{pv79} in Appendix~\ref{app-cmunurho}.


\section{Form-factor decompositions of the helicity amplitudes}
\label{sec-amplitudes}
\cleqn

Eqns.~(\ref{eq-brs1}), (\ref{eq-brs2}) and (\ref{doublebrs}) present
identities among amplitudes as derived from BRS symmetry.  One of our
goals is to rewrite these identities as sum rules among scalar form factors.
Hence, in this section we present the most general form-factor decompositions
of the various amplitudes, and we introduce important notation and terminology.


\subsection{$e^+e^- \rightarrow W^+W^-$}
\label{subsec-amp-eeww}

\begin{figure}[tbhp]
\begin{center}
\leavevmode\psfig{file=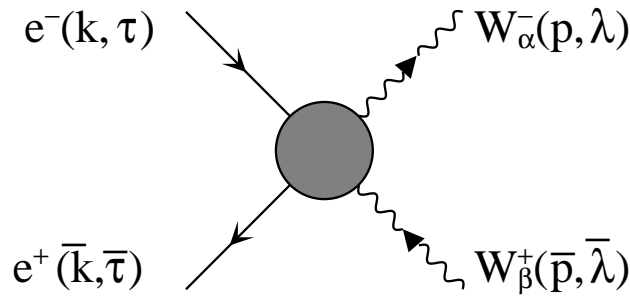,angle=0,height=4cm,silent=0}
\end{center}
\caption{The process $e^-e^+ \rightarrow W^-W^+$\/ with momentum and helicity 
assignments.  The momenta $k$ and $\overline{k}$ are incoming, but $p$ and 
$\overline{p}$ are outgoing.  The arrows on the $W$-boson lines indicate the 
flow of negative electric charge.}
\label{fig-eeww-blob} 
\end{figure}

The process 
\begin{equation}
\label{eeww}
e^-(k,\tau) + e^+(\overline{k},\overline{\tau}) \rightarrow
W^-(p,\lambda) + W^+(\overline{p},\overline{\lambda})
\end{equation}
is depicted in  Fig.\ref{fig-eeww-blob}.
The four-momenta of the $e^-$, $e^+$, $W^-$ and $W^+$ are $k$, $\overline{k}$,
$p$ and $\overline{p}$, respectively.  The helicity of the $e^-$ ($e^+$) is 
given by $\frac{1}{2}\tau$ ($\frac{1}{2}\overline{\tau}$), and $\lambda$
($\overline{\lambda}$) is the helicity of the $W^-$ ($W^+$) boson.  In the 
limit of
massless electrons only $\overline{\tau} = -\tau$ amplitudes survive, and the 
most general amplitude for this process may be written as \cite{hhis97}
\begin{equation}
\label{amp-eeww}
{\cal M}_{e^-e^+\rightarrow W^-W^+}
(k,\overline{k},\tau;p,\overline{p},\lambda,\overline{\lambda})
= \sum_{i=1}^{16} F_{i,\tau}(s,t)\, j_\mu(k,\overline{k},\tau) 
T_i^{\mu\alpha\beta} \epsilon_\alpha(p,\lambda)^\ast 
\epsilon_\beta(\overline{p},\overline{\lambda})^\ast  \;,
\end{equation}
where all dynamical information is contained in the scalar form-factors 
$F_{i,\tau}(s,t)$ with $s = (k + \overline{k})^2$ and $t = (k - p)^2$.  
The remaining factors on the right-hand side of Eqn.~(\ref{amp-eeww}) are 
purely kinematical; $\epsilon_\alpha(p,\lambda)^\ast$ and 
$\epsilon_\beta(\overline{p},\overline{\lambda})^\ast$ are the polarization 
vectors for the $W^-$ and $W^+$ bosons, respectively, and 
$j_\mu(k,\overline{k},\tau)$ is the massless electron current,
\begin{equation}
\label{fcurrent}
 j_\mu(k,\overline{k},\tau) = \overline{v}(\overline{k},-\tau) 
\gamma_\mu u(k,\tau) \;.
\end{equation}

The summation over $i$ merits further discussion.  The tensors, 
$T_i^{\mu\alpha\beta}$, have three independent Lorentz indices, and each 
Lorentz index may assume four values.  Hence, one might expect that 
there are $4^3 = 64$ independent form factors.  One factor of four corresponds 
to the degrees of freedom for the $W^-$ boson: three physical 
polarizations plus one unphysical degree of freedom.  We refer to the physical 
polarizations as the transverse and longitudinal polarizations, and the 
unphysical degree of freedom is the scalar polarization.  Likewise, for the 
$W^+$ boson there are three physical polarizations plus one unphysical scalar 
polarization.  Finally, there are two ways to align plus two ways to anti-align
the electron and positron helicities yielding the third factor of four.
However, since we are working in the limit of massless electrons, we only 
need to consider the two helicity-aligned states, implying 32 independent 
form factors.  We also find that one set of tensors may be used for both 
the left- and right-handed electron currents.  Hence, the sum runs from 
$i=1$ to 16, and each form factor carries a subscript $\tau$.  

The three physical polarization vectors are orthogonal to the momentum of the 
$W$ boson,
\begin{equation}
p_\mu \epsilon^\mu(p,\lambda)^\ast = 
\overline{p}_\mu \epsilon^\mu(\overline{p},\overline{\lambda})^\ast = 0\;,
\label{eq-physpol}
\end{equation}
and the scalar polarization is defined proportional to momentum vector by
\begin{equation}
\epsilon^\mu(p,S)^\ast = \frac{p^\mu}{m_W}\;, \makebox[0.5cm]{}
\epsilon^\mu(\overline{p},S)^\ast = \frac{\overline{p}^\mu}{m_W}\;, 
\label{eq-scalpol}
\end{equation}
where $m_W$ is the physical mass of the $W$ boson.
Hence the 
appearance of factors such as $p^\alpha$ and $\overline{p}^\beta$ in a tensor 
$T_i^{\mu\alpha\beta}$ indicate that its associated form factor contributes 
only for unphysical amplitudes.  Furthermore, factors of $q^\mu$
do not survive due to current conservation, and factors of $k^\mu$ or 
$\overline{k}^\mu$ don't survive due to the Dirac equation for massless 
fermions, \ie\/ $k\!\!\!/u(k) = 0$ and 
$\overline{v}(\overline{k})\overline{k}\!\!\!/ = 0$.  Then, the sixteen 
tensors may be chosen as
\begin{subequations}
\begin{eqnarray}
\label{ff_T1}
T_1^{\mu\alpha\beta} & = &  P^\mu g^{\alpha\beta} \;,\\
\label{ff_T2}
T_2^{\mu\alpha\beta} & = & \frac{-1}{m_W^2} P^\mu q^\alpha q^\beta \;,\\
\label{ff_T3}
T_3^{\mu\alpha\beta} & = & q^\alpha g^{\mu\beta} - q^\beta g^{\alpha\mu} \;,\\
\label{ff_T4}
T_4^{\mu\alpha\beta} & = & 
               i \Big( q^\alpha g^{\mu\beta} + q^\beta g^{\alpha\mu} \Big) \;,\\
\label{ff_T5}
T_5^{\mu\alpha\beta} & = & i \epsilon^{\mu\alpha\beta\rho} P_\rho \;,\\
\label{ff_T6}
T_6^{\mu\alpha\beta} & = & - \epsilon^{\mu\alpha\beta\rho} q_\rho \;,\\
\label{ff_T7}
T_7^{\mu\alpha\beta} & = & \frac{-1}{m_W^2} P^\mu 
                    \epsilon^{\alpha\beta\rho\sigma} q_\rho P_\sigma \;,\\
\label{ff_T8}
T_8^{\mu\alpha\beta} & = &  K^\beta g^{\alpha\mu} + K^\alpha g^{\mu\beta}\;,\\
\label{ff_T9}
T_9^{\mu\alpha\beta} & = &  \frac{i}{m_W^2} 
                            \Big( K^\alpha \epsilon^{\beta\mu\rho\sigma} 
           + K^\beta \epsilon^{\alpha\mu\rho\sigma} \Big) q_\rho P_\sigma\;,\\
\label{ff_T10}
T_{10}^{\mu\alpha\beta} & = &  p^\alpha g^{\mu\beta}\;,\\
\label{ff_T11}
T_{11}^{\mu\alpha\beta} & = &  \frac{1}{\mwsq}P^\mu P^\beta p^\alpha \;,\\
\label{ff_T12}
T_{12}^{\mu\alpha\beta} & = &  \frac{i}{\mwsq}
                               \epsilon^{\mu\beta\rho\sigma}
                               P_\rho q_\sigma p^\alpha \;,\\
\label{ff_T13}
T_{13}^{\mu\alpha\beta} & = &  \overline{p}^\beta g^{\mu\alpha}\;,\\
\label{ff_T14}
T_{14}^{\mu\alpha\beta} & = &  \frac{1}{\mwsq}P^\mu P^\alpha 
                               \overline{p}^\beta \;,\\
\label{ff_T15}
T_{15}^{\mu\alpha\beta} & = &  \frac{i}{\mwsq}
                               \epsilon^{\mu\alpha\rho\sigma}
                               P_\rho q_\sigma \overline{p}^\beta \;,\\
\label{ff_T16}
T_{16}^{\mu\alpha\beta} & = & \frac{1}{\mwsq}P^\mu p^\alpha \overline{p}^\beta
                              \;,
\end{eqnarray}
\label{ff_T1-ff_T16}
\end{subequations}
where
\begin{subequations}
\label{tensor-variables}
\begin{eqnarray}
P & = & p-\overline{p} \;, \\ 
q & = & k+\overline{k} = p+\overline{p} \;,\\ 
K & = & k-\overline{k} \;,
\end{eqnarray}
\end{subequations}
and 
$\epsilon_{0123} = -\epsilon^{0123} = +1$.  We explicitly use the {\em 
physical}
$W$-boson mass, $m_W$, in the definition of the tensors; this choice will be
important when discussing one-loop corrections.  Tensors 
$T_1^{\mu\alpha\beta}$ through $T_9^{\mu\alpha\beta}$, which were listed in 
Ref.~\cite{hhis97}, are sufficient to describe all physical amplitudes.   
The first seven tensors are sufficient to describe the most general 
$\gamma WW$ and $Z WW$ vertex functions \cite{hpzh87}.  Tensors 
$T_{10}^{\mu\alpha\beta}$ through $T_{12}^{\mu\alpha\beta}$ contribute if 
the $W^-$ boson has a scalar polarization, $T_{13}^{\mu\alpha\beta}$ through 
$T_{15}^{\mu\alpha\beta}$ contribute if the $W^+$ boson is unphysical, 
and $T_{16}^{\mu\alpha\beta}$ is needed only when both $W$ bosons have scalar 
polarizations.


\subsection{$e^+e^-\rightarrow \chi^+W^-$ and $e^+e^-\rightarrow W^+\chi^-$}
\label{subsec-amp-eewx}

\begin{figure}[tbhp]
\begin{center}
\leavevmode\psfig{file=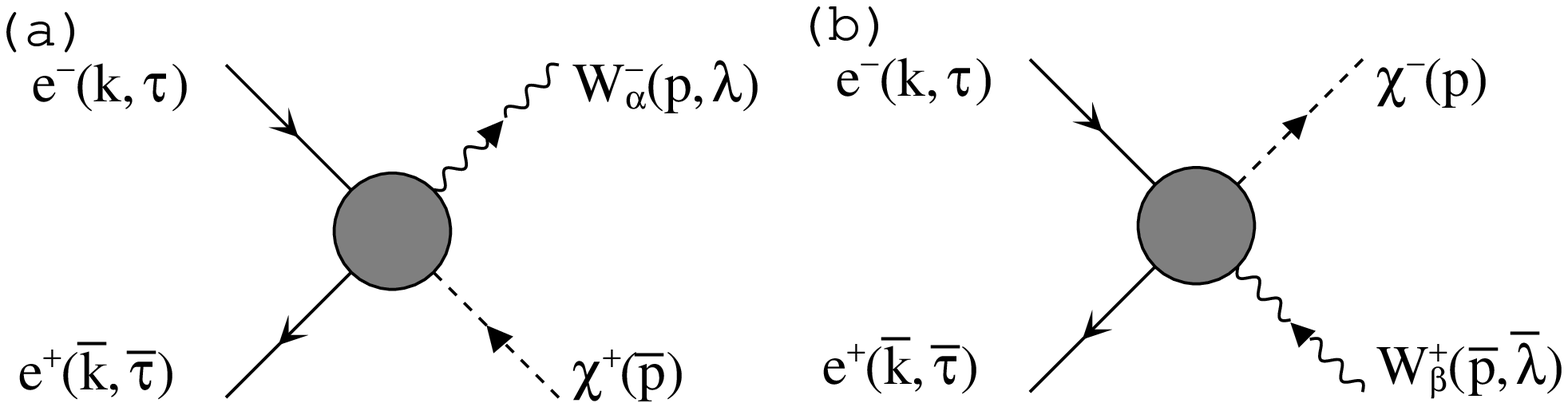,angle=0,height=4cm,silent=0}
\end{center}
\caption{The processes (a) $e^+e^-\rightarrow W^-\chi^+$ and (b)
$e^+e^-\rightarrow \chi^-W^+$ with momentum and helicity assignments 
chosen to coincide as closely as possible to those in 
Fig.~\protect{\ref{fig-eeww-blob}}. The momenta $k$ and $\overline{k}$ are 
incoming, but $p$ and $\overline{p}$ are outgoing.  The arrows along the 
$W$-boson and Goldstone-boson lines indicate the flow of negative electric
charge.}
\label{fig-eewx-eexw-blob}
\end{figure}

Next we consider amplitudes for $e^+e^- \rightarrow W^\pm \chi^\mp$.  See 
Figs.~\ref{fig-eewx-eexw-blob}(a) and \ref{fig-eewx-eexw-blob}(b).
Our phase convention for the Goldstone bosons is that of Ref.~\cite{hisz93}.
We decompose the amplitudes as follows:
\begin{subequations}
\begin{eqnarray}
 {\cal M}_{e^-e^+\rightarrow W^-\chi^+}
(k,\overline{k},\tau;p,\overline{p},\lambda)
& = &   i \sum_{i=1}^{4} H_{i,\tau}(s,t)\, j_\mu(k,\overline{k},\tau) 
S_i^{\mu\alpha} \epsilon_\alpha(p,\lambda)^\ast \;,\label{amp-eewx}\\
 {\cal M}_{e^-e^+ \rightarrow \chi^- W^+}
(k,\overline{k},\tau;p,\overline{p},\overline{\lambda})
& = &  i \sum_{i=1}^{4} \overline{H}_{i,\tau}(s,t) 
 \, j_\mu(k,\overline{k},\tau) \overline{S}_i^{\mu\beta} 
\epsilon_\beta (\overline{p},\overline{\lambda})^\ast \;.\label{amp-eexw}
\end{eqnarray}
\end{subequations}
There are four independent tensors, $S_i^{\mu\alpha}$, corresponding to the 
four polarizations, three physical plus one scalar, of the $W^-$ boson in 
$W^-\chi^+$ production, and each form factor $H_{i,\tau}(s,t)$ carries an 
index for the electron helicity, $\tau$.  The tensors are given by
\begin{subequations}
\label{ff_Si}
\begin{eqnarray}
\label{ff_S1}
S_1^{\mu\alpha} & = & m_W g^{\mu\alpha} \;,\\
\label{ff_S2}
S_2^{\mu\alpha} & = & \frac{1}{m_W}P^\mu P^\alpha \;, \\
\label{ff_S3}
S_3^{\mu\alpha} & = & \frac{i}{m_W} \epsilon^{\mu\alpha\rho\sigma} 
P_\rho q_\sigma \;,\\
\label{ff_S4}
S_4^{\mu\alpha} & = & \frac{1}{m_W} P^\mu p^\alpha \;,
\end{eqnarray}
\end{subequations}
where $P^\mu$ and $q^\mu$ are given by Eqns.~(\ref{tensor-variables}).
The tensors $S_1^{\mu\alpha}$ through $S_3^{\mu\alpha}$ contribute to 
$W^-\chi^+$ production when the $W^-$ has a  physical polarization, while 
$S_4^{\mu\alpha}$ contributes for the scalar polarization.  A second set of 
four tensors is introduced for $W^+\chi^-$ production:
\begin{subequations}
\label{ff_barSi}
\begin{eqnarray}
\label{ff_barS1-S3}
\overline{S}_i^{\mu\beta} & = & S_i^{\mu\beta} \makebox[.6cm]{} 
(i = 1,2,3) \;,\\
\label{ff_barS4}
\overline{S}_4^{\mu\beta} & = & \frac{1}{m_W} P^\mu \overline{p}^\beta\;.
\end{eqnarray}
\end{subequations}
$\overline{S}_1^{\mu\beta}$ through $\overline{S}_3^{\mu\beta}$ contribute 
for physical polarizations, and $\overline{S}_4^{\mu\beta}$ contributes 
for the scalar polarization of the $W^+$ boson.  The corresponding 
form factors are given by $\overline{H}_{i,\tau}(s,t)$.


\subsection{$e^+e^- \rightarrow \chi^+ \chi^-$}
\label{subsec-amp-eexx}

\begin{figure}[tbhp]
\begin{center}
\leavevmode\psfig{file=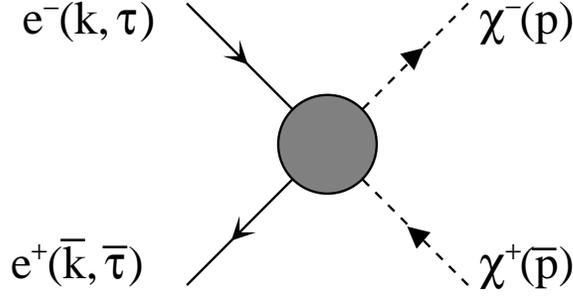,angle=0,height=4cm,silent=0}
\end{center}
\caption{The process $e^+e^-\rightarrow \chi^-\chi^+$ with momentum and 
helicity assignments chosen to coincide as closely as possible to those in 
Fig.~\protect{\ref{fig-eeww-blob}}. The momenta $k$ and $\overline{k}$ are 
incoming, but $p$ and $\overline{p}$ are outgoing.  The arrows along the 
Goldstone-boson lines indicate the flow of negative electric charge.}
\label{fig-eexx-blob}
\end{figure}

The process $e^+e^- \rightarrow \chi^+\chi^-$ is shown in 
Fig.~\ref{fig-eexx-blob}.
The corresponding amplitude may be decomposed as 
\begin{equation}
{\cal M}(e^+e^- \rightarrow \chi^+\chi^-) 
       =  P^\mu j_\mu(k,\bar{k},\tau) R_\tau(s,t)\;.
\label{amp-wwxx}       
\end{equation}
Notice that there is only one form factor, $R_\tau(s,t)$, which carries an
index for the electron helicity.


\section{Applications of BRS Symmetry}
\label{sec-brs}
\cleqn

Now that we have presented the form-factor decompositions for our diboson 
production amplitudes we can recast our single BRS identities of 
Eqns.~(\ref{eq-brs1}) and (\ref{eq-brs2}) and our double BRS identity of 
Eqn.~(\ref{doublebrs}) as sum rules among the form factors.  In this section 
we will also comment on the relationship between our exact sum rules and the 
equivalence theorem which only holds in the high-energy limit.  

Before we proceed it is convenient to present the relationships between the 
rank-three tensors, $T_i^{\mu\alpha\beta}$, the rank-two tensors, 
$S_i^{\mu\alpha}$ and $\overline{S}_i^{\mu\beta}$, and the rank-one tensor, 
$P^\mu$, which were used in the decompositions of $e^+e^- \rightarrow W^+W^-$, 
$e^+e^- \rightarrow \chi^\pm W^\mp$ and $e^+e^- \rightarrow \chi^+\chi^-$
amplitudes.  In general we can write 
\begin{subequations}
\begin{eqnarray}
   T_i^{\mu\alpha\beta} \epsilon_\beta (\bar{p},S)^\ast & = & 
  \sum_{j=1}^{4}  \xi_{ij}(s,t) S_j^{\mu\alpha} \;,
\label{eq-exp1}  \\
   T_i^{\mu\alpha\beta} \epsilon_\alpha (p,S)^\ast  & = & 
  \sum_{j=1}^{4} \overline{\xi}_{ij}(s,t)  \overline{S}_j^{\mu\beta} \;,
\label{eq-exp2}  
\end{eqnarray}
\end{subequations}
and 
\begin{subequations}
\begin{eqnarray}
\label{eq-etadef}
  S_{j}^{\mu\alpha} \epsilon_\alpha (p,S)^\ast & = & \eta_j P^\mu\;, \\
  \overline{S}_{j}^{\mu\beta} 
   \epsilon_\beta (\overline{p},S)^\ast & = & \overline{\eta}_j P^\mu\;.  
\label{eq-etabardef}
\end{eqnarray}
\end{subequations}
The coefficients $\xi_{ij}(s,t)$ and $\overline{\xi}_{ij}(s,t)$,
which are dimensionless functions of the Mandelstam variables $s$ and $t$,
are given in Table~\ref{table-xi}. 
In Table~\ref{table-eta} we list the coefficients $\eta_j$ and 
$\overline{\eta}_j$. The expressions 
\begin{table}[htbp]
\begin{center}
\begin{tabular}{|r||cccc|}\hline
$i$     & $\xi_{i1}$ & $\xi_{i2}$ & $\xi_{i3}$ & $\xi_{i4}$ \\ \hline
$1$     &      & $-1$  &                  & $1$    \\ 
$2$     &      &  $2 \gamma^2$ &        & $- 4 \gamma^2$ \\ 
$3$     & $- 2 \gamma^2$ & $1/2$  &   & $- 1$  \\ 
$4$     & $i 2 \gamma^2$ & $i/2$ &    & $- i$  \\ 
$5$     &   &   & $-1/2$                  &  \\ 
$6$     &   &   & $-i/2$ &  \\ 
$7$     &   &   &   &  \\ 
$8$     & $4 \delta^2$ &   & $-\tau/2$ &  \\ 
$9$     &   &   & $2 \delta^2$ &  \\    
$10$    &   &   &   &    $ - 1/2 $ \\ 
$11$    &   &   &   &    $ 2 \gamma^2 - 2 $  \\ 
$12$    &   &   &   &      \\ 
$13$    & $1$ &   &   &   \\ 
$14$    &   & $1$ &   &   \\ 
$15$    &   &   & $1$ &   \\  
$16$    &   &   &   &   $1$ \\ \hline 
\end{tabular}
\hspace{1cm}
\begin{tabular}{|r||cccc|}\hline
$i$ &  $\overline{\xi}_{i1}$ &$\overline{\xi}_{i2}$ &$\overline{\xi}_{i3}$ 
       &$\overline{\xi}_{i4}$ \\ \hline
$i$     &    & $1$  &                 &   $1$    \\ 
$2$     &      & $- 2 \gamma^2$ &           & $- 4 \gamma^2$ \\ 
$3$     & $2 \gamma^2$ & $- 1/2$  &       & $-1$\\ 
$4$     & $i 2 \gamma^2$ & $i/2$ &       & $ i$  \\ 
$5$     &   &   & $1/2$                  &  \\ 
$6$     &   &   & $-i/2$ &  \\ 
$7$     &   &   &   &  \\ 
$8$     &  $- 4\delta^2$ &   & $\tau/2$ &  \\ 
$9$     &   &   & $- 2 \delta^2$ &  \\   
$10$    & $1$ &   &   &    \\ 
$11$    &   & $1$ &   &      \\ 
$12$    &   &   & $1$ &      \\ 
$13$    &   &   &   & $ 1/2 $ \\ 
$14$    &   &   &   & $ - 2\gamma^2 + 2 $ \\ 
$15$    &   &   &   &  \\  
$16$    &   &   &   & $ 1 $\\ \hline 
\end{tabular}
\caption{The coefficients $\xi_{ij}(s,t)$ of Eqn.~(\protect{\ref{eq-exp1}})
and $\overline{\xi}_{ij}(s,t)$ of Eqn.~(\protect{\ref{eq-exp2}}).  Only 
nonzero values are shown.}
\label{table-xi}
\end{center}
\end{table}
\begin{table}[htbp]
\begin{center}
\begin{tabular}{|r||cccc|}\hline
         &  $j=1$ & $j=2$ & $j=3$   & $j=4$ \\ \hline\hline
$\eta_j$ & $1/2$ & $-2(\gamma^2 - 1)$ & $0$ & $1$   \\ 
$\overline{\eta}_j$ & $-1/2$ & $2(\gamma^2 - 1)$ & $0$ & $1$ \\ \hline  
\end{tabular}
\caption{The coefficients $\eta_j$ and $\overline{\eta}_j$ of 
Eqns.~(\protect{\ref{eq-etadef}}) and (\protect{\ref{eq-etabardef}}),
respectively.}
\label{table-eta}
\end{center}
\end{table}
%
\begin{subequations}
\begin{eqnarray}
  \gamma^2 & = & \frac{s}{4 m_W^2}\;, \label{gamma}\\ 
  \delta^2 & = & \frac{s + 2 t - 2 m_W^2}{4 m_W^2}\;, \label{delta}
\end{eqnarray}
\end{subequations}
are used in the tables.  


\subsection{`Single' BRS sum rules}
\label{subsec-single}

Because we frequently refer to the combinations of amplitudes which appear in
the single BRS identities of Eqns.~(\ref{eq-brs1}) and (\ref{eq-brs2}),
it is convenient to define $G_{i,\tau}$ and $\overline{G}_{i,\tau}$ by
\begin{subequations}
\label{eq-g-and-barg}
\begin{eqnarray}
\label{eq-g}
j_\mu(k,\overline{k},\tau) G_{j,\tau} S_j^{\mu\alpha} 
\epsilon_\alpha(p,\lambda)^\ast & = & 
{\cal M}(e^+e^-\rightarrow W^-_P W^+_S) + 
i\, C_{\rm mod}{\cal M}(e^+e^-\rightarrow W^-_P \chi^+)    \;,\\
\label{eq-barg}
j_\mu(k,\overline{k},\tau) \overline{G}_{j,\tau}
\overline{S}_j^{\mu\beta} 
\epsilon_\beta(\overline{p},\overline{\lambda})^\ast
& = & {\cal M}(e^+e^-\rightarrow W^-_S W^+_P) + 
i\,C_{\rm mod} {\cal M}(e^+e^-\rightarrow \chi^- W^+_P)\;.
\end{eqnarray}
\end{subequations}
BRS symmetry then implies $G_{i,\tau} = \overline{G}_{i,\tau} = 0$.
Next, we rewrite Eqn.~(\ref{amp-eeww}) requiring that the $W^+$ boson has an 
unphysical scalar polarization and the $W^-$ is physical.  Employing the 
expansion of Eqn.~(\ref{eq-exp1}) we obtain
\begin{equation}
{\cal M} (e^+e^- \rightarrow W^-_PW^+_S)
 = \sum_{j=1}^3 
  \left\{ \sum_{i=1}^{16} \xi_{ij}(s,t) F_{i,\tau} (s,t) \right\} 
  j_\mu (k,\bar{k},\tau) S_j^{\mu\alpha} \epsilon_\alpha (p,P)^\ast \;,
\label{eq-exlhs1}
\end{equation}
where $S_4^{\mu\alpha}$ does not contribute by Eqn.~(\ref{eq-physpol}).  
If we insert this expression along with the expansion of the 
$e^+e^-\rightarrow W^-\chi^+$ amplitude of Eqn.~(\ref{amp-eewx}) into 
Eqn.~(\ref{eq-brs1}), then we find the following sum rules: 
\begin{equation}
  G_{j,\tau}(s,t) =
  \sum_{i=1}^{16}  F_{i,\tau}(s,t) \xi_{i j}(s,t) 
  - C_{\rm mod} H_{j,\tau}(s,t)
   = 0\;,
\label{eq-brssum1}
\end{equation}
for $j = 1, 2, 3$.
In a parallel fashion, if we rewrite Eqn.~(\ref{amp-eeww}) for a physical 
$W^+$ boson and an unphysical $W^-$, then applying Eqn.~(\ref{eq-exp2}) yields
\begin{equation}
{\cal M} (e^+e^- \rightarrow W^-_SW^+_P)
 = \sum_{j=1}^3 \left\{ \sum_{i=1}^{16} \overline{\xi}_{ij}(s,t) 
  F_{i,\tau} (s,t) \right\} 
  j_\mu (k,\bar{k},\tau) \overline{S}_j^{\mu\beta} 
  \epsilon_\beta (\overline{p},P)^\ast .
\label{eq-exlhs2}
\end{equation}
We insert this in Eqn.~(\ref{eq-brs2}) along with Eqn.~(\ref{amp-eexw})
to obtain
\begin{equation}
  \overline{G}_{j,\tau}(s,t) = \sum_{i=1}^{16} 
  F_{i,\tau}(s,t) \overline{\xi}_{ij}(s,t)
  - C_{\rm mod} \overline{H}_{j,\tau}(s,t) 
   = 0\;,
\label{eq-brssum2}  
\end{equation}
for $j = 1, 2, 3$.  

Inserting the coefficients $\xi_{ij}$ as listed in Table~\ref{table-xi} we 
write
\begin{subequations}
\begin{eqnarray}
\!\!\!\!  -2 \gamma^2 \,
  \Big\{ F_{3,\tau}(s,t) - i\, F_{4,\tau}(s,t) \Big\}
  + 4 \delta^2\, F_{8,\tau}(s,t) +  F_{13,\tau}(s,t) 
  - C_{\rm mod} H_{1,\tau}(s,t)  &=& 0\;, \;\;
\label{eq-expbrs1a}
\\  
\!\!\!\!  - F_{1,\tau}(s,t) 
  + 2 \gamma^2\, F_{2,\tau}(s,t) 
  + \frac{1}{2} F_{3,\tau}(s,t) 
  + \frac{i}{2} F_{4,\tau}(s,t) 
  + F_{14,\tau}(s,t) 
  - C_{\rm mod} H_{2,\tau}(s,t)  &=& 0\;, \;\;
\label{eq-expbrs1b}
\\ 
\!\!\!\!  - \frac{1}{2} F_{5,\tau}(s,t) 
  - \frac{i}{2} F_{6,\tau}(s,t)
  - \frac{\tau}{2} F_{8,\tau}(s,t)
  + 2 \delta^2\, F_{9,\tau}(s,t)
  + F_{15,\tau}(s,t) 
  - C_{\rm mod} H_{3,\tau}(s,t) &=& 0\;, \;\;
\label{eq-expbrs1c}
\end{eqnarray}
\end{subequations}
corresponding to $G_{1,\tau} = 0$, $G_{2,\tau} = 0$ and $G_{3,\tau} = 0$,
respectively.  Taking the $\overline{\xi}_{ij}$ from Table~\ref{table-xi} 
produces
\begin{subequations}
\begin{eqnarray}
\!\!\!\!
    2 \gamma^2 \Big\{ F_{3,\tau}(s,t) +  i\, F_{4,\tau}(s,t) \Big\}
  - 4 \delta^2\, F_{8,\tau}(s,t) 
  +  F_{10,\tau}(s,t)  
  - C_{\rm mod} \overline{H}_{1,\tau}(s,t) &=& 0\;, \;\;
\label{eq-expbrs2a}
\\  
\!\!\!\!
    F_{1,\tau}(s,t) 
  - 2 \gamma^2\, F_{2,\tau}(s,t) 
  - \frac{1}{2} F_{3,\tau}(s,t) 
  + \frac{i}{2} F_{4,\tau}(s,t) 
  + F_{11,\tau}(s,t) 
  - C_{\rm mod} \overline{H}_{2,\tau}(s,t) &=& 0\;, \;\;
\label{eq-expbrs2b}
\\ 
\!\!\!\!
  \frac{1}{2} F_{5,\tau}(s,t) 
  - \frac{i}{2} F_{6,\tau}(s,t)
  + \frac{\tau}{2} F_{8,\tau}(s,t)
  - 2 \delta^2\, F_{9,\tau}(s,t)
  + F_{12,\tau}(s,t)
  - C_{\rm mod} \overline{H}_{3,\tau}(s,t) &=& 0\;.\;\;
\label{eq-expbrs2c}
\end{eqnarray}
\end{subequations}
corresponding to $\overline{G}_{j,\tau} = 0$ for $j = 1, 2, 3$.
Note that the expression of these sum rules holds for each order 
of the perturbative expansion. 
Eqns.~(\ref{eq-expbrs1a})-(\ref{eq-expbrs1c}), taking into account 
the two electron helicity states, are a set of six sum rules.  Of the 
eighteen form factors that contribute to physical $W$-boson pair production 
amplitudes, {\em i.e.} $F_{1,\tau}$ through $F_{9,\tau}$, all but the 
CP-violating $F_{7,\tau}$ appear. ($F_{7,\tau}$ only contributes to final 
states where both $W$ bosons have the same transverse polarization, hence 
cannot contribute to these sum rules.)  In order to test {\em via}
Eqns.~(\ref{eq-expbrs1a})-(\ref{eq-expbrs1c}) the remaining sixteen,
we must also retain $F_{10,\tau}$ through $F_{12,\tau}$ in the evaluation of 
the $e^+e^-\rightarrow W^+W^-$ amplitudes. It is furthermore necessary to 
obtain $H_{1,\tau}$ through $H_{3,\tau}$ from the independent evaluation of 
the $e^+e^-\rightarrow \chi^+W^-$ amplitudes.  The test is extremely powerful 
and hence worthy of the extra effort.  Each form factor can have its own 
complicated dependence on $s$ and $t$.  Each can include a large number of 
loop diagrams. Yet when added together they satisfy the above sum rules.

Eqns.~(\ref{eq-expbrs2a})-(\ref{eq-expbrs2c}) yield an additional set of 
six sum rules which again test all of the physically relevant form factors 
except $F_{7,\tau}$.  In this case we must retain $F_{13,\tau}$ through 
$F_{15,\tau}$ in the evaluation of the $e^+e^-\rightarrow W^+W^-$ amplitudes
while obtaining $\overline{H}_{1,\tau}$ through $\overline{H}_{3,\tau}$ from 
the evaluation of the $e^+e^-\rightarrow W^+\chi^-$ amplitudes.
As a practical matter, once one set of sum rules, either 
Eqns.~(\ref{eq-expbrs1a})-(\ref{eq-expbrs1c}) or 
Eqns.~(\ref{eq-expbrs2a})-(\ref{eq-expbrs2c}), is used to verify the 
accuracy of a calculation, the other set is redundant.


\subsection{`Double' BRS Sum rules}
\label{subsec-double}

Next we rewrite Eqn.~(\ref{amp-eeww}) requiring that both $W$ bosons have 
unphysical scalar polarizations.  Then we can use Eqn.~(\ref{eq-exp1}) 
followed by Eqn.~(\ref{eq-etadef}), or we can use Eqn.~(\ref{eq-exp2}) 
followed by Eqn.~(\ref{eq-etabardef}) to obtain
\begin{eqnarray}
  {\cal M}(e^+e^- \rightarrow W^+_SW^-_S) 
     &=& P^\mu j_\mu(k,\bar{k},\tau)
         \sum_{j=1}^4 \sum_{i=1}^{16} F_{i,\tau}(s,t) \xi_{ij}(s,t) \eta_j 
   \nonumber \\
&=&  P^\mu j_\mu(k,\bar{k},\tau)
         \sum_{j=1}^4 \sum_{i=1}^{16} F_{i,\tau}(s,t) 
       \bar{\xi}_{ij}(s,t) \bar{\eta}_j \;. \label{wsws}
\end{eqnarray}
Then, from Eqns.~(\ref{amp-eewx}) and (\ref{amp-eexw}), we have
\begin{subequations}
\begin{eqnarray}
   {\cal M} (e^+e^- \rightarrow \chi^+ W^-_S) 
  &=& i 
      P^\mu j_\mu(k,\bar{k},\tau) \sum_j^4 H_{j,\tau}(s,t)\, \eta_j
  \;,  \\
   {\cal M} (e^+e^- \rightarrow W^+_S \chi^-) 
  &=& i  P^\mu j_\mu(k,\bar{k},\tau) \sum_j^4 
    \overline{H}_{j,\tau}(s,t)\, \overline{\eta}_j
   \;. 
\end{eqnarray}
\end{subequations}
Incorporating the expansion of the $e^+e^-\rightarrow \chi^+\chi^-$ 
amplitude in Eqn.~(\ref{amp-wwxx}), the double BRS identity of 
Eqn.~(\ref{doublebrs})  becomes
\begin{subequations}
\begin{eqnarray}
\!\!\!\!\!\!\!\!\!\!\!\!\!\!\!\!\!\!\!\!
  \sum_{i=1}^{16} F_{i,\tau}(s,t) \xi_{ij}(s,t)
                \eta_j  
  - C_{\rm mod} \sum_{j=1}^4 H_{j,\tau}(s,t) \eta_j 
  - C_{\rm mod} \sum_{j=1}^4 \overline{H}_{j,\tau}(s,t) \overline{\eta}_j 
 - \left\{ C_{\rm mod} \right\}^2 R_\tau (s,t) & = & 0\;,  \label{eq-wbrs1} \\
\!\!\!\!\!\!\!\!\!\!\!\!\!\!\!\!\!\!\!\
  \sum_{i=1}^{16} F_{i,\tau}(s,t) \overline{\xi}_{ij}(s,t)
                \overline{\eta}_j 
  - C_{\rm mod} \sum_{j=1}^4 H_{j,\tau}(s,t) \eta_j 
  - C_{\rm mod} \sum_{j=1}^4 \overline{H}_{j,\tau}(s,t) \overline{\eta}_j 
  - \left\{ C_{\rm mod} \right\}^2  R_\tau (s,t) & = & 0\;.  \label{eq-wbrs2} 
\end{eqnarray}
\end{subequations}


\subsection{Relationship to the equivalence theorem}
\label{subsec-et}

Here, we would like to comment on the relationship between the BRS sum rules 
and the equivalence theorem\cite{et,et2,et3}.  The theorem states that the 
amplitudes for a process with longitudinally polarized gauge bosons $W_L^\pm$ 
are equivalent at energies where $s \gg m_W^2$ to those of the process in 
which the $W_L^\pm$ bosons are replaced by their corresponding Goldstone 
bosons.  The theorem follows from BRS symmetry and the 
equivalence of the longitudinal and scalar polarization vectors at 
high energies.  Consider $W$-boson pair production in the center of momentum 
frame with the $W^-$ boson momentum along the positive $z$ axis.  In this frame
the longitudinal and scalar polarization vectors are given 
by
\begin{subequations}
\begin{eqnarray}
\epsilon^{\mu} (p,0)^\ast & = & \gamma \Big( \beta , 0 , 0 , 1 \Big) \;, \\
\epsilon^{\mu} (p,S)^\ast & = & \gamma \Big( 1 , 0 , 0 , \beta \Big) \;, \\
\epsilon^{\mu} (\overline{p},0)^\ast & = & \gamma \Big( \beta , 0 , 0 , 
        -1 \Big) \;,\\
\epsilon^{\mu} (\overline{p},S)^\ast & = & \gamma \Big( 1 , 0 , 0 , 
        -\beta \Big) \;.
\end{eqnarray}
\end{subequations}
Thus for energies large compared to the $W$-boson mass we have  
\begin{subequations}
\begin{eqnarray}
\epsilon^{\mu} (p,0)^\ast  
           & = &  \epsilon^{\mu} (p,S)^\ast
          \left( 1 + {\cal O}(m_W/\sqrt{s}) \right)\;, 
           \label{polarization1}\\
\epsilon^{\mu} (\overline{p},0)^\ast  
           & = &  \epsilon^{\mu} (\overline{p},S)^\ast 
          \left( 1 + {\cal O}(m_W/\sqrt{s}) \right)\;. 
\label{polarization2}
\end{eqnarray}
\end{subequations}
Hence we may rewrite the `single' BRS identities of Eqns.~(\ref{eq-brs1}) 
and (\ref{eq-brs2}) to find
\begin{subequations}
\begin{eqnarray}
{\cal M}\Big(e^+ e^- \rightarrow W^-_L\, W^+_L\Big) 
&=& - i C_{\rm mod} {\cal M}\Big(e^+ e^- \rightarrow W^-_L\, \chi^+\Big) 
\, \left( 1 +  {\cal O}(m_W/\sqrt{s})  \right)\;, \label{et1} \\
{\cal M}\Big(e^+ e^- \rightarrow W^-_L\, W^+_L\Big) 
&=& - i C_{\rm mod} {\cal M}\Big(e^+ e^- \rightarrow \chi^- \,W^+_L\Big) \, 
 \left( 1 +  {\cal O}(m_W/\sqrt{s}) \right)\;, \label{et2}
\end{eqnarray}
\end{subequations}
while, for the double BRS identity of Eqn.~({\ref{doublebrs}), we obtain 
\begin{eqnarray}&&
{\cal M} (e^+e^- \rightarrow W^+_L\,W^-_L) 
 + i C_{\rm mod} {\cal M} (e^+e^- \rightarrow W^+_L\, \chi^-) \nonumber\\&&
\makebox[0.5cm]{}
 + i C_{\rm mod} {\cal M} (e^+e^- \rightarrow \chi^+ \,W^-_L) 
 = \left\{ C_{\rm mod} \right\}^2  
{\cal M} (e^+e^- \rightarrow \chi^+ \,\chi^-) \,
\left( 1 +  {\cal O}(m_W/\sqrt{s}) \right)\;.  
\label{et3}
\end{eqnarray}
Inserting Eqns.~(\ref{et1}) and (\ref{et2}) into Eqn.~(\ref{et3}), we obtain  
\begin{eqnarray}
  {\cal M}\Big(e^+ e^- \rightarrow W^-_L\, W^+_L\Big) 
&=& \left\{- i C_{\rm mod} \right\}^2 
 {\cal M}\Big(e^+ e^- \rightarrow \chi^- \, \chi^+\Big)\,
\left( 1 +  {\cal O}\left(m_W/\sqrt{s}\right) \right)\;. \label{et4}
\end{eqnarray}
Although the equivalence theorem has often been used to check the consistency
of various calculations (for example, see Ref.~\cite{sk97}), it is valid only 
in the high-energy limit. On the other hand, our BRS sum rules hold exactly at 
all energies.   Therefore, where perturbation theory is applicable, BRS sum 
rules are a more powerful tool.


\section{Calculation of the form factors in perturbation theory}
\label{sec-ff}
\cleqn

In later sections we will present detailed one-loop calculations 
concerning 
$e^+e^- \rightarrow W^+W^-$ and $e^+e^- \rightarrow W^\pm\chi^\mp$ 
amplitudes.  We will then check the calculation by using the BRS sum rules.
Hence, in this section we present a discussion of the related
form factors.

It is extremely important that we avoid any inconsistent treatment of 
higher order effects. 
Because we want to employ the physical $W$-boson mass 
we choose $m_W$ as one of our input parameters. 
We also choose  $\hat{e}$ and $\hat{s}$, the 
${\overline{\rm MS}}$ parameters of the electric coupling constant and 
sine of the Weinberg angle, respectively. 
These three parameters are consistently employed in the evaluation 
of all loop integrals and form factors. 
The ${\overline{\rm MS}}$ weak coupling constants $\hat{g}$ and 
$\hat{g}_Z$ are related to  $\hat{e}$ and $\hat{s}$ by 
\begin{eqnarray}
  \hat{c}^2 = 1 - \hat{s}^2,\;\; 
  \hat{e} = \hat{g} \hat{s} = \hat{g}_Z \hat{s} \hat{c}\; .  
\end{eqnarray}
The ${\overline{\rm MS}}$ masses of the massive gauge bosons 
are defined in terms of $\hat{e}$, $\hat{s}$ and $m_W$ by 
\begin{eqnarray}
  \hat{m}_W^2 &=& m_W^2 + \Pi_T^{WW}(m_W^2) \;,\\
  \hat{m}_Z^2 &=& \frac{\hat{m}_W^2}{\hat{c}^2} 
   = \frac{1}{\hat{c}^2} 
     \left\{ m_W^2 + \Pi_T^{WW} (m_W^2) \right\}\; ,
\end{eqnarray}
and the physical mass of the $Z$ boson is obtained (at the one-loop level) by  
\begin{eqnarray}
  m_Z^2 = \frac{m_W^2}{\hat{c}^2} 
   + \frac{1}{\hat{c}^2} \Pi_T^{WW} (m_W^2) - \Pi_T^{ZZ} 
 \left(\frac{m_W^2}{\hat{c}^2}\right) \equiv 
  \frac{m_W^2}{\hat{c}^2} + \Delta\; , \label{defmz}
\end{eqnarray}
where the contributions of the two-point functions are represented by 
$\Delta$.   
Finally, when $m_Z^2$ appears in the denominator 
we perform an expansion and truncate after the term which is linear in 
$\Delta$ as 
\begin{eqnarray}
  \frac{1}{s - m_Z^2} = \frac{1}{s - (m_W^2/\hat{c}^2)} 
     \left\{ 1 + \frac{\Delta}{s - m_W^2/\hat{c}^2}\right\}\;. \label{expmz}
\end{eqnarray}
In Ref.~\cite{achksu}, we will demonstrate that, by following this scheme, 
we are able to satisfy the BRS sum rules to the limit of floating-point 
prescription.


\subsection{$e^+e^- \rightarrow W^+W^-$}
\label{subsec-ff-eeww}

The form factors $F_{i,\tau}(s,t)$ which were introduced in 
Eqn.~(\ref{amp-eeww}) may be written as
\begin{eqnarray}
\nonumber
&&\makebox[-1cm]{} 
F_{i,\tau}(s,t)  =  \frac{\hatesq}{s}\Bigg\{\bigg[Q_e \Big(
        1 - \pitggg(s) + \Gamma_1^{\,e}(s)\Big) + T^3_e\, 
        \overline{\Gamma}\mbox{}^{\,e}_2(s) 
        \bigg]f_i^{\gamma\,(0)} 
        +  Q_e f_i^{\gamma\,(1)}(s) \Bigg\}
\\ \nonumber && 
\makebox[-0.5cm]{} 
        + \frac{\hatgsq}{s - (m_W^2/\hat{c}^2)}
        \Bigg\{\bigg[\big(T^3_e-\hatssq Q_e\big)
        \Big( 1 + 
             \frac{\Delta}{s - m_W^2/\hat{c}^2}
 - \pitzzz(s) + \Gamma_1^{\,e}(s) \Big) 
        + T^3_e \Big( \hat{c}^2 \overline{\Gamma}_2^{\,e}(s)  
+   \Gamma_3^{\,e}(s) \Big) \\ \nonumber&& 
\makebox[1.6cm]{}  
+ \Gamma_4^{\,e}(s) \bigg]f_i^{Z\,(0)} 
        + \big(T^3_e-\hatssq Q_e\big) f_i^{Z\,(1)}(s)
        - \frac{\hats}{\hatc} \bigg[ Q_e \hatcsq f_i^{Z\,(0)} 
        + \big(T^3_e-\hatssq Q_e\big) f_i^{\gamma\,(0)} \bigg] \pitgzg(s)
           \Bigg\}
\\ && 
\makebox[-0.5cm]{} 
        + \frac{T^3_e\hatgsq}{2t} 
\Big[ f_i^{t\,(0)} + \Gamma^{e\nu W}(t) + \overline{\Gamma}^{e\nu W}(t) \Big] 
        + F_{i,\tau}^{[\rm Box]}(s,t)\;,
\label{big-F}
\end{eqnarray}
where the electric charge and the third component of weak isospin of the 
electron are given by $Q_e = -1$ and $T^3_e$, respectively.  For left-handed
electrons $T^3_e = -\frac{1}{2}$, and $T^3_e=0$ for right-handed electrons.
We have already expanded the $Z$-boson propagator according to 
Eqn.~(\ref{expmz}).
The other factors in Eqn.~(\ref{big-F}) are explained in the text below.

The tree-level $e^+e^- \rightarrow W^+W^-$ Feynman diagrams are shown 
in Fig.~\ref{fig-eeww-tree}.  
\begin{figure}[tbhp]
\begin{center}
\leavevmode\psfig{file=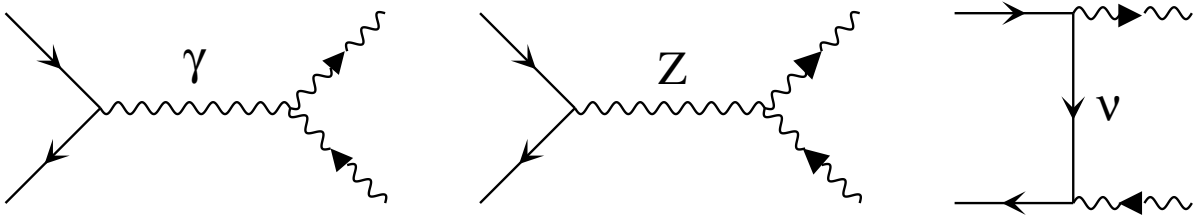,angle=0,height=3cm,silent=0}
\end{center}
\caption{Tree-level Feynman diagrams for $e^+e^- \rightarrow W^+W^-$.
The arrows on the $W$-boson lines indicate the flow of negative electric 
charge}
\label{fig-eeww-tree}
\end{figure}
Neglecting loop contributions the poles in $1/s$, $1/(s-\mzsq)$ and $1/t$ in 
Eqn.~(\ref{big-F}) correspond to the $s$-channel photon exchange, 
$s$-channel $Z$-boson exchange and $t$-channel neutrino exchange graphs,
respectively.  The appearance of the $\pitgzg$ term indicates $\gamma-Z$
mixing in the neutral current (NC) sector.  In general we define
\begin{equation}
\label{defn-pitv}
\Pi^{V_1V_2}_{T,V_3}(\qsq) = 
\frac{\Pi^{V_1V_2}_T(\qsq) - \Pi^{V_1V_2}_T(m^2_{V_3})}  {\qsq - m^2_{V_3}}\;,
\end{equation}
 which reduces to a derivative in the limit $m^2_{V_3}\rightarrow q^2$. 
Such an expression appears naturally when expanding a full propagator about 
the pole in the transverse part of the one-particle irreducible propagator
function.

Our Feynman rules for the $WW\gamma$ and $WWZ$ vertices are written as 
\begin{equation}
-i\hate\sum_{i = 1}^{16} f^\gamma_i T_i^{\mu\alpha\beta}
\makebox[3cm]{\rm and} 
-i\hatgz\hatcsq\sum_{i = 1}^{16} f^Z_i T_i^{\mu\alpha\beta}\;,
\end{equation}
respectively.  This introduces the vertex form factors $f^\gamma_i$ and 
$f^Z_i$.  The $t$-channel term is also expanded in the same 
$T_i^{\mu\alpha\beta}$ tensors\footnote{When casting the SM tree-level amplitude in the form of 
          Eqn.~(\ref{big-F})  we encountered a tensor 
          $T^{\mu\alpha\beta}_{17} = i \epsilon^{\mu\alpha\beta\rho}K_\rho$ 
          that contributes when the fermion current is left-handed. 
          An explicit calculation showed that $j_\mu(k,\overline{k},\tau) 
          T^{\mu\alpha\beta}_{17}
          \epsilon_\alpha(p,\lambda)^\ast 
          \epsilon_\beta(\overline{p},\overline{\lambda})^\ast 
          = \tau j_\mu(k,\overline{k},\tau) T^{\mu\alpha\beta}_3 
          \epsilon_\alpha(p,\lambda)^\ast 
          \epsilon_\beta(\overline{p},\overline{\lambda})^\ast $ for all 
          polarization choices, and hence it may be eliminated in lieu of 
          $T^{\mu\alpha\beta}_{3}$.}  
 yielding the form factors $f^t_i$. 
In turn 
each of these form factors is split into its tree-level value plus a one-loop 
correction as
\begin{equation}
f_i^V = f_i^{V\,(0)} + f_i^{V\,(1)}\;,
\end{equation}
where $V = \gamma$ and $Z$.  The nonzero tree-level values are given in 
Table~\ref{table-smf0}.
\begin{table}[tbh]
\begin{center}
\begin{tabular}{|c||cccccccccccccccc|}\hline
i & 1 & 2 & 3 & 4 & 5 & 6 & 7 & 8 & 9 & 10 & 11 & 12 & 13 & 14 & 15 & 16 \\ 
\hline \hline
$f_{i}^{\gamma\,(0)}$ &1&&2&& &&& &&$-1$&&&1&&& \\ 
$f_{i}^{Z\,(0)}$      &1&&2&& &&& &&$-1$&&&1&&& \\ 
$f_{i}^{t\,(0)}$      &1&&2&&1&&&1&&$-2$&&&2&&& \\ \hline
\end{tabular}
\end{center}
\caption{Explicit values for the $f_i^{X\,(0)}$ form factors of the SM at the 
tree level.  Only nonzero values are shown.}
\label{table-smf0}
\end{table}
Please note that $T_8^{\mu\alpha\beta}$ and $T_9^{\mu\alpha\beta}$ are 
precluded from contributing to the vector-boson vertex functions by angular 
momentum considerations\cite{hisz93}.  One-loop corrections to the $WW\gamma$ 
vertex are then contained in the $f_i^{\gamma\,(1)}$ along with the $Z$ factors
for the external $W$ bosons, $Z_W^{1/2}$, which multiply the corresponding 
$f_i^{\gamma\,(0)}$.  Similarly, one-loop corrections to the $WWZ$ vertex along
with the $Z_W^{1/2}$ factors that multiply the tree-level $WWZ$ vertex are 
contained in the $f_i^{Z(1)}$.  
The vertex functions for the $Vee$ vertex, denoted by 
$\Gamma_1^{\,e}$, $\overline{\Gamma}\mbox{}^{\,e}_2$, $\Gamma_3^{\,e}$ and 
$\Gamma_4^{\,e}$ also appear in 
$e^+e^-\rightarrow f\overline{f}$ amplitudes\cite{hhkm94}; 
the corrections from the external electron two-point functions are 
absorbed into $\Gamma_1$.  
The vertex functions 
$\Gamma^{\,e\nu W}$ and $\overline{\Gamma}\mbox{}^{\,e\nu W}$ 
appear in charged current processes; they contain $\nu e W$ vertex corrections 
as well as two-point function corrections 
for the external electrons and $W$ bosons and the internal neutrino 
propagator. 
Finally, the 
$F_{i,\tau}^{[\rm Box]}$ terms account for contributions of box diagrams.


\subsection{$e^+e^- \rightarrow \chi^+W^-$ and $e^+e^- \rightarrow W^+ \chi^-$}
\label{subsec-ff-eewx}

We are not interested, {\em per se}, in the form factors $H_{i,\tau}(s,t)$ of 
Eqn.~(\ref{amp-eewx}) and $\overline{H}_{i,\tau}(s,t)$ of 
Eqn.~(\ref{amp-eexw}). 
Rather we are interested in 
$C_{\rm mod} \HHbar_{i,\tau}(s,t)$ 
which is the expression that appears in the sum rules. 
Because $Z_W^{1/2} Z_\chi^{1/2} = 1$, 
Eqn.~(\ref{cmod}) simplifies to  
\begin{eqnarray}
  C_{\rm mod} = Z_m = \frac{\hat{m}_W}{m_W}\;. \label{cmod2}
\end{eqnarray}
Then we may compactly write
\begin{eqnarray}
\nonumber
 C_{\rm mod} \lefteqn{\HHbar\mbox{}_{i,\tau}(s,t) = 
\frac{\hatesq}{s}\Bigg\{\bigg[ Q_e \Big( \frac{\hat{m}_W^2}{m_W^2} 
- \pitggg(s) + \Gamma_1^{\,e}(s)
\Big) + T^3_e \overline{\Gamma}\mbox{}^{\,e}_2(s) \bigg]
\hhbar\mbox{}_i^{\gamma\,(0)} 
+ Q_e \hhbar\mbox{}_i^{\gamma\,(1)}(s) \Bigg\}}
&&\\
\nonumber && \makebox[-1.5cm]{} 
+ \frac{\hatgsq}{s  - \mwsq/\hat{c}^2}
\Bigg\{ \bigg[ (T^3_e-\hatssq Q_e)\Big( 
\frac{\hat{m}_W^2}{m_W^2} + 
\frac{\Delta}{s-m_W^2/\hat{c}^2}
  - \pitzzz(s) + \Gamma_1^{\,e}(s)
\Big) 
+ T^3_e \Big( \hat{c}^2 \overline{\Gamma}_2^{\,e}(s) + \Gamma_3^{\,e}(s) \Big)
\nonumber \\&&
\makebox[-1.5cm]{}
+ \Gamma_4^{\,e}(s)\bigg] \hhbar\mbox{}_i^{Z\,(0)}  + 
(T^3_e-\hatssq Q_e) \hhbar\mbox{}_i^{Z\,(1)}(s)  
- \frac{\hats}{\hatc} 
\bigg[ (T^3_e-\hatssq Q_e)\hhbar\mbox{}_i^{\gamma\,(0)} 
+ Q_e \hat{c}^2 \hhbar\mbox{}_i^{Z\,(0)} \bigg]
\pitgzg(s) \Bigg\} + \HHbar\mbox{}_{i,\tau}^{\,\boxes}(s,t) \;.\nonumber\\
\label{big-H}
\end{eqnarray}
At the tree level there are two Feynman graphs as shown 
in Fig.~\ref{fig-eewx-tree}; 
\begin{figure}[tbhp]
\begin{center}
\leavevmode\psfig{file=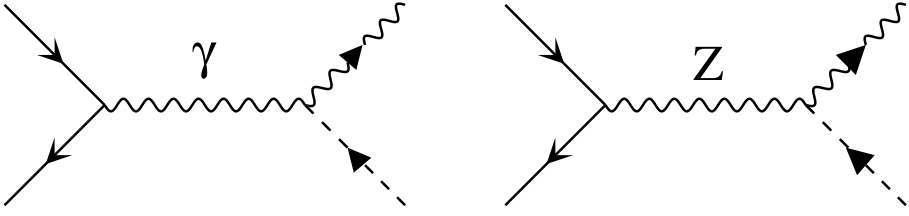,angle=0,height=3cm,silent=0}
\end{center}
\caption{Tree-level Feynman diagrams contributing to $e^+e^- \rightarrow 
W^-\chi^+$.  The arrows on the charged bosons indicate the flow of negative 
electric charge.  By reversing the direction of these arrows we obtain the 
graphs for $e^+e^- \rightarrow W^+\chi^-$.}
\label{fig-eewx-tree}
\end{figure}
in the limit of massless electrons there is no $t$-channel graph.  
The discussion of the hatted couplings, the pole mass and the two-point 
function corrections follow those of the previous section.  The 
expansion of $H_{i,\tau}(s,t)$ introduces the vertex form factors $h_i^\gamma$ 
and $h_i^Z$,  while the expansion of $\overline{H}_{i,\tau}(s,t)$ introduces 
$\overline{h}_i^\gamma$ and $\overline{h}_i^Z$.  The corresponding 
$VW^\mp\chi^\pm$ vertices are then expressed as
\begin{eqnarray}
\hate \sum_{i = 1}^4 h_i^{\gamma} S_i^{\mu\alpha}\;,\makebox[1cm]{}
\hatgz\hatcsq \sum_{i = 1}^4 h_i^{Z} S_i^{\mu\alpha}\;,\makebox[1cm]{}
\hate \sum_{i = 1}^4 \overline{h}_i^{\,\gamma} \overline{S}_i^{\mu\beta}\;,
\makebox[1cm]{} \hatgz\hatcsq \sum_{i = 1}^4 \overline{h}_i^{Z} 
\overline{S}_i^{\mu\beta}\;.
\end{eqnarray}
We use the physical 
mass in defining our tensors while the $VW^\pm\chi^\mp$ coupling depends 
explicitly on $\hat{m}_W$.  As before, the vertex form factors are written as 
the sum of the tree-level and one-loop contributions by
\begin{equation}
\hhbar\mbox{}_i^V(s) = \hhbar\mbox{}_i^{V\,(0)}\frac{\hat{m}_W}{m_W} 
+ \hhbar\mbox{}_i^{V\,(1)}\;,
\end{equation}
for $V = \gamma$, $Z$. 
This explains the appearance of one power of the ratio of the 
$\overline{\rm MS}$ $W$-boson mass to the physical mass 
multiplying the leading-order term. 
The other power comes from $C_{\rm mod}$ as in Eqn.~(\ref{cmod2}). 
The tree-level form-factor coefficients may be found in 
Table~\ref{table-smhhbar0}.
The $\hhbar\mbox{}_i^{V\,(1)}$ contain not only vertex corrections for the 
$VW\chi$ vertices but also the correction factors for the external 
$W$ bosons.
\begin{table}[tbh]
\begin{center}
\begin{tabular}{|c||cccc|c|c||cccc|}\cline{1-5}\cline{7-11}
$\makebox[1cm]{i}$ & $\makebox[1cm]{1}$ & $\makebox[1cm]{2}$ & 
$\makebox[1cm]{3}$ & $\makebox[1cm]{4}$ & $\makebox[1cm]{}$
& $\makebox[1cm]{i}$ & $\makebox[1cm]{1}$ & $\makebox[1cm]{2}$ & 
$\makebox[1cm]{3}$ & $\makebox[1cm]{4}$ \\\cline{1-5} \cline{7-11}
$h_i^{\gamma\,(0)}$ & $1$ & &&&
& $\overline{h}_i^{\,\gamma\,(0)}$ & $-1$ &&& \\
$h_i^{Z\,(0)}$ & $-\hatssq/\hatcsq$ & &&&& $\overline{h}_i^{Z\,(0)}$ & 
$\hatssq/\hatcsq$ &&& \\
\cline{1-5} \cline{7-11}
\end{tabular}
\end{center}
\caption{Explicit values for the nonzero $h_i^{V\,(0)}$ and 
$\overline{h}_i^{V\,(0)}$ form factors of the SM at the tree level.}
\label{table-smhhbar0}
\end{table}


\section{BRS Sum Rules at the tree level}
\label{sec-tree}
\cleqn

Before we embark on the study of the BRS sum rules at the loop level it is 
instructive to first study them at the tree level.  We first 
consider the sum rule in Eqn.~(\ref{eq-expbrs2a}).  We obtain the tree-level
form factors from Eqns.~(\ref{big-F}) and (\ref{big-H}) by retaining only the 
Born term and setting $m_Z = \hat{m}_Z$ and $m_W = \hat{m}_W$.  Then,
\begin{eqnarray}
\nonumber
\overline{G}_{1,\tau} & = &  2\gamma^2\bigg\{ 2\frac{Q_e\hatesq}{s} 
+ 2\frac{(T^3_e-\hatssq Q_e)\hatgzsq\hatcsq}{s-\hat{m}_Z^2} 
+ 2 \frac{T^3_e \hatgsq}{2t}\bigg\} - 4\delta^2 \bigg\{ 
\frac{T^3_e\hatgsq}{2t} \bigg\}
\\ && \mbox{}
 - \bigg\{ \frac{Q_e\hatesq}{s} 
+ \frac{(T^3_e-\hatssq Q_e)\hatgzsq\hatcsq}{s-\hat{m}_Z^2} 
+ 2 \frac{T^3_e \hatgsq}{2t}\bigg\}
-  \bigg\{ - \frac{Q_e\hatesq}{s} + 
\frac{(T^3_e-\hatssq Q_e)\hatgzsq\hatssq}{s-\hat{m}_Z^2}\bigg\}\;.
\end{eqnarray}
Reading from left to right the four terms correspond to the contributions 
of $F_{3,\tau}$, $F_{8,\tau}$, $F_{10,\tau}$ and $\overline{H}_{1,\tau}$; 
$F_{4,\tau}$ also contributes but is identically zero at the tree-level.  
See Table~\ref{table-smf0}.  Regrouping the terms by electron quantum 
numbers yields
\begin{eqnarray}
\nonumber
\overline{G}_{1,\tau} & = & Q_e\hatesq\Bigg\{ 4\gamma^2 
\bigg[ \frac{1}{s} - \frac{1}{s-\hat{m}_Z^2} \bigg] 
- \bigg[ \frac{1}{s} - \frac{1}{s-\hat{m}_Z^2} \bigg] 
+ \bigg[ \frac{1}{s} + \frac{1}{s-\hat{m}_Z^2}\frac{\hatssq}{\hatcsq} \bigg] 
\Bigg\}\\
&&\mbox{} + T^3_e\hatgsq \Bigg\{\Big(4\gamma^2-4\delta^2 - 2\Big)\frac{1}{2t}
+ \Big(4\gamma^2 - 1 - \frac{\hatssq}{\hatcsq} \Big)\frac{1}{s-\hat{m}_Z^2}
\Bigg\}\;.
\end{eqnarray}
The coefficient of the pole in $1/t$ comes entirely from the neutrino-exchange
diagram in $W$-boson pair production.  We notice that 
$-4\gamma^2+4\delta^2 + 2 = 2t/\hat{m}_W^2$, and hence the dependence on $t$ 
cancels between the various $W$-pair contributions. After some simplification 
we find
\begin{eqnarray}
\overline{G}_{1,\tau} & = & - \Big\{Q_e\hatesq-T^3_e\hatgsq\Big\}
\frac{\hatcsq\hat{m}_Z^2-\hat{m}_W^2}{\hatcsq\hat{m}_W^2(s-\hat{m}_Z^2)}\;,
\label{gtree}
\end{eqnarray}
which vanishes because the tree level (or $\overline{\rm MS}$) couplings and 
masses satisfy
\begin{equation}
\label{weinberg}
\hat{m}_W^2 = \hatcsq\hat{m}_Z^2\;. \label{rel}
\end{equation}
The necessity of retaining the tree-level relation of Eqn.~(\ref{rel}) when
testing BRS invariance at the tree-level has been noted by the authors of
the HELAS\cite{helas} Fortran routines, and it is built into the default 
HELAS parameter set.  It is instructive to note that, in models with 
nondoublet, nonsinglet vacuum expectation values both Eqn.~(\ref{rel}) and 
the Goldstone-boson couplings are modified such that the BRS 
identities remain valid.

We could repeat this exercise for $\overline{G}_{1,\tau}$ to obtain at each 
stage, up to the overall sign, exactly the same result.  The other sum 
rules are trivial at the tree level.  Since $\sum f_i^{X\,(0)} \xi_{i2}
= \sum f_i^{X\,(0)} \overline{\xi}_{i2} = 0$ and $h^{V\,(0)}_2 = 
\overline{h}^{V\,(0)}_2 = 0$, the coefficients of the $s$, $s- \hat{m}_Z^2$ 
and 
$t$ poles are separately zero, hence $G_{2,\tau} = \overline{G}_{2,\tau} = 0$. 
We most easily obtain $G_{3,+} = \overline{G}_{3,+} = 0$ since every term is 
identically zero, while $G_{3,-} = \overline{G}_{3,-} = 0$ follows because 
$f_5^{t\,(0)}-f_8^{t\,(0)}=0$.  See 
Tables~\ref{table-xi}-\ref{table-smf0}.


\section{BRS Sum Rules at the one-loop level}
\label{sec-brs-loop}
\cleqn

The BRS invariance of the Lagrangian is a general result. Hence it may be 
applied order by order in perturbation theory.  While already 
useful at the tree level, we are interested in applying BRS sum rules to 
one-loop calculations.  We will find that our ensuing discussions are 
facilitated if we subdivide our form factors and sum rules according to 
the type of diagrams which contribute.  In some cases this division is a 
mere convenience which allows us to break large expressions into smaller ones.
But, in other cases, we find that the contributions are naturally divided
into BRS-invariant subsets.
 
We begin by writing the form factors at the one-loop level 
for $W$-boson pair production in the following form:
\begin{equation}
F_{i,\tau} =          F_{i,\tau}^\born
                  +   F_{i,\tau}^\nc
                  +   F_{i,\tau}^\gvtx
                  +   F_{i,\tau}^\zvtx
                  +   F_{i,\tau}^\wphys
                  +   F_{i,\tau}^\unphys
                  +   F_{i,\tau}^{[eeV]}
                  +   F_{i,\tau}^{[\nu e W]}
                  +   F_{i,\tau}^\boxes\;,
\label{eq-fdec1}
\end{equation}
where $F_{i,\tau}^\born$ contains all of the terms which remain when we
drop the two- and three-point function corrections and box terms.  It looks 
like the tree-level expression except the internal propagators are expressed
in terms of the physical masses.  The two-point-function corrections for the 
neutral-current $s$-channel contributions are in $F_{i,\tau}^{[NC]}$.  We have 
combined the three-point loop corrections with the $W$-boson wave-function 
renormalization factors to obtain the $f^{V\,(1)}_i$ form factors; this yields 
a finite result when the $W$ bosons are physical.  While this 
approach is ideal for the calculation of the physical amplitudes, here we 
find it more convenient to separate the vertex and external corrections.  The 
former are included in $F_{i,\tau}^\gvtx$ and $F_{i,\tau}^\zvtx$ while the 
latter are in $F_{i,\tau}^\wphys$.  Corrections for the external unphysical 
$W$ bosons and Goldstone bosons are the sole contributers to 
$F_{i,\tau}^\unphys$.  
Finally, the $\Gamma_1^e$, $\overline{\Gamma}_2^e$, $\Gamma_3^e$ and 
$\Gamma_4^e$ terms are combined in $F_{i,\tau}^{[eeV]}$, the 
$\Gamma^{\nu e W}$ and $\overline{\Gamma}^{\nu e W}$ terms comprise 
$F_{i,\tau}^{[\nu e W]}$, and the box terms form $F_{i,\tau}^{[\rm Box]}$.
We perform a parallel decomposition of the form factors for 
$W^\pm\chi^\mp$ production as
\begin{subequations}
\begin{eqnarray}
\label{eq-hdec1}
\HHbar_{i,\tau}  =       \HHbar_{i,\tau}^\born
                  +   \HHbar_{i,\tau}^\nc
                  +   \HHbar_{i,\tau}^\gvtx
                  +   \HHbar_{i,\tau}^\zvtx
                  +   \HHbar_{i,\tau}^\wphys
                  +   \HHbar_{i,\tau}^\unphys
                  +   \HHbar_{i,\tau}^{[eeV]}
                  +   \HHbar_{i,\tau}^\boxes\;.\label{eq-hdec2}
\end{eqnarray}
\end{subequations}
Note that there is no $H_{i,\tau}^{[\nu e W]}$ term.  
In a natural way we can then rewrite 
the single BRS sum rules of  Eqns.~(\ref{eq-brssum1}) and (\ref{eq-brssum2}) 
as
\begin{eqnarray}
\GGbar_{j,\tau} 
  =   \sum_{X} 
    \bigg\{ \sum^{16}_{i=1} \xixibar_{i j} F_{i,\tau}^{[X]} 
   -    C_{\rm mod}
\HHbar_{j,\tau}^{[X]} \bigg\}  = 
    \sum_X \GGbar_{j,\tau}^{[X]} = 0\;,
  \label{eq-defgx2}   
\end{eqnarray}
where $[X]$ generically denotes any of the contributions discussed above.
Now we may proceed by considering terms of the form $G_{i,\tau}^{[X]}$
and $\overline{G}_{i,\tau}^{[X]}$ one at a time.

We first consider the contributions of wave-function renormalization factors 
for the physical $W$ bosons.  These factors merely multiply the purely 
tree-level amplitudes.  We have already shown the BRS invariance of the purely 
tree-level amplitudes, and hence we know that
\begin{subequations}
\begin{eqnarray}
  \GGbar^\wphys_{i,\tau} &=& 0\;.\label{wf2}
\end{eqnarray}
\end{subequations}
This is actually unfortunate because it means that, since this type of 
correction only changes the overall normalization, our BRS sum rules do not 
test the correction factors for the physical $W$ bosons.  

By virtue of the Ward-Takahashi identities for the two-point functions of the 
unphysical bosons\cite{lv80},
\begin{eqnarray}
  \Pi^{{W_S}{W_S}}(q^2) + i \Pi^{\chi{W_S}}(q^2) + i \Pi^{{W_S}\chi}(q^2)
     - \Pi^{\chi\chi}(q^2) = 0\;,
\end{eqnarray}
we can show that
\begin{eqnarray}
  \GGbar^\unphys_{i,\tau} &=& 0 \;.\label{unphys2}
\end{eqnarray}
The explicit proof of Eqn.~(\ref{unphys2}) is given in 
Appendix~\ref{app-unphysical}.  This time we have found a welcome 
simplification.  Our main interest is to test the physically relevant form
factors, $F_{1,\tau}(s,t)$ through $F_{9,\tau}(s,t)$.  The contributions 
from the unphysical two-point functions form a set which independently 
satisfies the BRS sum rules.  Hence we can, for all practical purposes, 
neglect these contributions.

Next we will show that 
\begin{eqnarray}
  \GGbar_{i,\tau}^\born + \GGbar_{i,\tau}^\nc
                          + \GGbar_{i,\tau}^\gvtx
                          + \GGbar_{i,\tau}^\zvtx = 0 \;.   
\label{brsin}
\end{eqnarray}
For clarity of presentation we will focus on $\overline{G}_{1,\tau}$, but 
we can easily extend the discussion to the other sum rules as well.  
The Born
term looks similar to the purely tree-level results, but there are 
a few differences.  
First, 
the $Z$-boson mass in the $Z$-boson propagator is 
the physical mass which must be calculated from the input parameters, 
$\hat{m}_W$, $\hat{e}$ and $\hat{s}$. 
The $Z$-boson propagator must then be expressed in $\Delta$ as in 
Eqn.~(\ref{defmz})   
Second, a factor of $\hat{m}_W^2/m_W^2$ multiplies 
the Born term in  the $\HHbar_{i,\tau}$ form factors 
(see Eqn.~(\ref{big-H})). Hence, 
\begin{eqnarray}
\nonumber
 \makebox[-3cm]{} \overline{G}_{1,\tau}^\born 
  &=& \frac{Q_e \hatesq}{s}\sum_i f_i^{\gamma\,(0)}
+ \frac{(T^3_e -\hatssq Q_e)\hatgsq}{s - (m_W^2/\hat{c}^2)} 
(1 + \frac{\Delta}{s - m_W^2/\hat{c}^2}) 
\sum_i f_i^{Z\,(0)}
+ \frac{T^3_e \hatgsq}{2t}\sum_i f_i^{t\,(0)} \nonumber\\
&&
\;\;\;\;\;\;\;\;
 + \bigg\{\frac{Q_e \hatesq}{s}\frac{\hat{m}_W^2}{\mwsq}
-  (\frac{\hat{m}_W^2}{\mwsq} + \frac{\Delta}{s - m_W^2/\hat{c}^2}) 
\frac{(T^3_e -\hatssq Q_e)\hatgsq}{s - (m_W^2/\hat{c}^2)}  
\frac{\hatssq}{\hatcsq}\bigg\} \nonumber\\ 
 \makebox[-3cm]{}
 &=& \frac{Q_e \hatesq}{s}\frac{s-\mwsq}{\mwsq}
 + \frac{(T^3_e -\hatssq Q_e)\hatgsq}{s-(m_W^2/\hat{c}^2)} 
  \frac{s-\mwsq}{\mwsq}(1 + \frac{\Delta}{s - m_W^2/\hat{c}^2})
- \frac{T^3_e\hatgsq}{\mwsq} \nonumber\\
&&+ \frac{Q_e \hatesq}{s}\frac{\hat{m}_W^2}{\mwsq}
- (\frac{\hat{m}_W^2}{\mwsq} + \frac{\Delta}{s - m_W^2/\hat{c}^2})  
\frac{(T^3_e -\hatssq Q_e)\hatgsq}{s-\mwsq/\hat{c}^2}
\frac{\hatssq}{\hatcsq} \;.
\end{eqnarray}
With some rearrangement we
are able to write the results as
\begin{eqnarray}
\overline{G}_{1,\tau}^\born & = & \frac{Q_e\hatesq}{s m_W^2}\pitww(\mwsq)
+ \frac{(T^3_e -\hatssq Q_e)\hatgsq}{(s-m_W^2/\hat{c}^2) {m}_W^2}
\bigg\{\pitww(\mwsq)-\pitzz(\mwsq/\hat{c}^2)\bigg\}\;.
\label{gborn}
\end{eqnarray}
The calculation of NC contributions is straightforward and leads to 
\begin{eqnarray}
\makebox[-.8cm]{}
\overline{G}_{1,\tau}^\nc & = & - \frac{Q_e\hatesq}{s {m}_W^2}
\bigg\{ \pitgg(s) + \frac{\hatc}{\hats}\pitgz(s) \bigg\}
- \frac{(T^3_e -\hatssq Q_e) \hatgsq}{(s- m_W^2/\hat{c}) {m}_W^2}
\bigg\{\pitzz(s)-\pitzz(\mwsq/\hat{c}^2)+\frac{\hats}{\hatc}\pitgz(s)\bigg\}\;.
\nonumber\\
\label{gnc}
\end{eqnarray}
To obtain Eqn.~(\ref{brsin}) we must now turn our attention
to the vertex corrections.  Our task is made much simpler by considering the 
following Ward identities\cite{lv80,fr97}:
\begin{subequations}
\begin{eqnarray}
\label{wtidgamma}
  p_\alpha \Gamma_{\mu\alpha\beta}^{\gamma WW}(s,p,\bar{p}) 
  + i {m}_W \Gamma_{\mu\beta}^{\gamma \chi W}(s,p,\bar{p})  
  & = & \hat{e}
  \left\{ \Pi_{\mu\beta}^{WW}(\bar{p}) - \Pi_{\mu\beta}^{\gamma\gamma}(q) 
  + \frac{\hat{c}}{\hat{s}}\Pi_{\mu\beta}^{\gamma Z}(q) \right\}\;, \\
  p_\alpha \Gamma_{\mu\alpha\beta}^{ZWW}(s,p,\bar{p}) 
  + i {m}_W \Gamma_{\mu\beta}^{Z\chi W}(s,p,\bar{p})  
  & = &  \hat{c}^2 \hat{g}_Z \left\{ \Pi_{\mu\beta}^{WW}(\bar{p}) - 
  \Pi_{\mu\beta}^{ZZ}(q) 
 +  \frac{\hat{s}}{\hat{c}}\Pi_{\mu\beta}^{\gamma Z}(q) \right\}\;. 
\label{wtidz}
\end{eqnarray}
\end{subequations}
From here we may determine
\begin{subequations}
\begin{eqnarray}
\label{ggVTX} 
  \overline{G}_{1,\tau}^\gvtx & = & \frac{Q_e\hat{e}^2}{s {m}_W^2} 
   \left\{ - \Pi_T^{WW}(m_W^2) + \Pi_T^{\gamma\gamma}(s)
   - \frac{\hat{c}}{\hat{s}} \Pi_T^{\gamma Z}(s) \right\}\;,\\
  \overline{G}_{1,\tau}^\zvtx & = & 
\frac{(T^3_e-\hat{s}^2 Q_e)\hat{c}^2\hat{g}_Z^2}{(s-{m}_W^2/\hat{c}^2){m}_W^2} 
      \left\{ - \Pi_T^{WW}(m_W^2) 
           + \Pi_T^{ZZ}(s)
           - \frac{\hat{s}}{\hat{c}} \Pi_T^{\gamma Z}(s) \right\}\;.
 \label{gzVTX}
\end{eqnarray}
\end{subequations}
We may sum $\overline{G}_{1,\tau}^\born$, $\overline{G}_{1,\tau}^\nc$,
$\overline{G}_{1,\tau}^\gvtx$ and $\overline{G}_{1,\tau}^\zvtx$ by inspection.
The $\pitzz(\mwsq/\hat{c}^2)$ terms cancel between $\overline{G}_{1,\tau}^\born$
and $\overline{G}_{1,\tau}^\nc$ while the $\pitww(\mwsq)$ terms cancel 
between $\overline{G}_{1,\tau}^\born$, $\overline{G}_{1,\tau}^\gvtx$ and 
$\overline{G}_{1,\tau}^\zvtx$.  The remaining terms cancel between the NC
and vertex contributions.   This completes the proof of Eqn.~(\ref{brsin}).
The case for $G_{j,\tau}$ in Eqn.~(\ref{brsin}) 
can be verified by a similar argument.

There are only three remaining terms in Eqn.~(\ref{eq-defgx2}) for which we
may infer
\begin{eqnarray}  
      \overline{G}_{i,\tau}^{[eeV]} 
    + \overline{G}_{i,\tau}^{[\nu e W]} 
    + \overline{G}_{i,\tau}^\boxes
&=& 0\;. 
  \label{brsin3} 
\end{eqnarray}
This concludes the discussion for $\overline{G}_{1,\tau}$.  

The discussion for $\overline{G}_{1,\tau}$ may be applied to $G_{1,\tau}$ up 
to the overall sign.  For $G_{2,\tau}$, $\overline{G}_{2,\tau}$, $G_{3,\tau}$ 
and $\overline{G}_{3,\tau}$,  the discussion further simplifies since the
NC and Born contributions drop out; see the comments in the last paragraph of 
Section~\ref{sec-tree}.  We may summarize these results by splitting the 
twelve BRS identities of Eqns.~(\ref{eq-brssum1}) and (\ref{eq-brssum2}) into 
the following forty-eight invariant subsets:
\begin{subequations}
\begin{eqnarray}
\label{wp-set}
&&G_{i,\tau}^\wphys = \overline{G}_{i,\tau}^\wphys = 0\;,\\
\label{unphysical-set}
&&G_{i,\tau}^\unphys = \overline{G}_{i,\tau}^\unphys = 0\;,\\
\label{useful-set-1}
&&G_{i,\tau}^\born + G_{i,\tau}^\nc + G_{i,\tau}^\gvtx + G_{i,\tau}^\zvtx
= \overline{G}_{i,\tau}^\born + \overline{G}_{i,\tau}^\nc 
+ \overline{G}_{i,\tau}^\gvtx + \overline{G}_{i,\tau}^\zvtx  = 0\;,\\
\label{useful-set-2}
&&  \overline{G}_{i,\tau}^{[eeV]} 
    + \overline{G}_{i,\tau}^{[\nu e W]} + G_{i,\tau}^\boxes = 
     \overline{G}_{i,\tau}^{[eeV]} 
    + \overline{G}_{i,\tau}^{[\nu e W]} 
    +  \overline{G}_{i,\tau}^\boxes  = 0\;.
\end{eqnarray}
\end{subequations} 
The twenty-four sum rules in Eqns.~(\ref{useful-set-1}) and 
(\ref{useful-set-2}) are useful for testing physical amplitudes.


\section{Scalar-fermion contributions at one loop}
\label{sec-brs-sfermion}
\cleqn

So far we have been quite general in the presentation and discussion of the
BRS sum rules.  In this section we will discuss a concrete example. In general
it is our goal to use BRS sum rules to aid in difficult calculations.  For 
example, in a series of papers\cite{achksu} we are studying the complete 
one-loop contributions of the Minimal Supersymmetric Standard Model 
(MSSM).  That is a large and complex calculation, so BRS sum rules will 
provide a valuable test of the results.  In this section, for the purpose of 
illustration, we will discuss a much simpler problem, the contributions of an 
SU(2)$_L$ doublet of scalar fermions, $(\tilde{u}_L, \tilde{d}_L)^T$, and 
their right-handed singlet counterparts, $\tilde{u}_R$ and $\tilde{d}_R$, to 
the $e^+e^-\rightarrow W^+W^-$ sum rules.  For simplicity left-right mixing 
is neglected.

Scalar-fermion corrections enter through corrections to the gauge-boson 
propagators as shown in Fig.~\ref{fig-vv}.
\begin{figure}[tbhp]
\begin{center}
\leavevmode\psfig{file=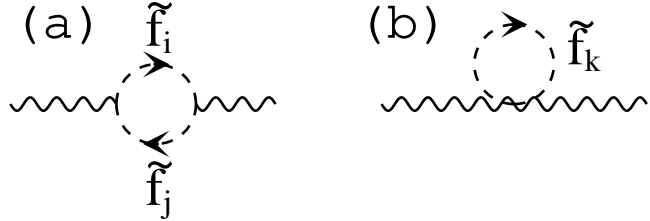,angle=0,height=3cm,silent=0}
\end{center}
\caption{Contributions of scalar fermions to the gauge-boson two-point
functions.  Both left- and right-handed scalar fermions contribute to $\pigg$, 
$\pigz$ and $\pizz$ while only the left-handed states may contribute to 
$\piww$.}
\label{fig-vv}
\end{figure}
The explicit formulae for the two-point-function corrections are given 
in Appendix~\ref{app-propagators}.  As demonstrated in the previous 
section, we can neglect two-point-function corrections for the external 
particles.  Appendix~\ref{app-unphysical} discusses the propagator 
corrections for the external $W_S$ and $\chi^\pm$ bosons which form a 
closed BRS-invariant subset.  Corrections for external physical $W$ bosons
are required to get the correct answer for physical amplitudes, but such
corrections only change the overall normalization and cannot be verified 
by these methods.

There are two categories of $WWV$ vertex corrections.  The first type, 
which we shall call scalar-fermion triangle (SFT) diagrams, are shown in 
Fig.~\ref{fig-wwv-triangle}.
\begin{figure}[tbhp]
\begin{center}
\leavevmode\psfig{file=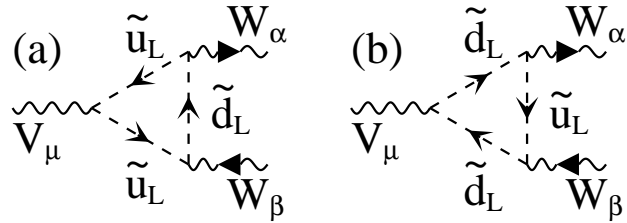,angle=0,height=3cm,silent=0}
\end{center}
\caption{Scalar-fermion triangle-type (SFT) one-loop corrections to the 
$WW\gamma$ and $WWZ$ vertices.  Because of the coupling to a $W$ boson, only 
the left-handed scalar fermions contribute.  The arrows on the $W$-boson lines 
indicate the flow of negative electric charge.}
\label{fig-wwv-triangle}
\end{figure}
These diagrams contribute to the $f_i^{\gamma\,(1)}$ and $f_i^{Z\,(1)}$ form 
factors of Eqn.~(\ref{big-F}) through the terms $f_i^{\gamma\,(1)\,{\rm SFT}}$ 
and $f_i^{Z\,(1)\,{\rm SFT}}$, respectively.  Explicit formulae are given in 
Appendix~\ref{app-vertices}.  The other type of vertex corrections employ one 
scalar-fermion seagull (SFSG) vertex and one three-point vertex.  They are 
shown in Fig.~\ref{fig-wwv-sg}.
\begin{figure}[tbhp]
\begin{center}
\leavevmode\psfig{file=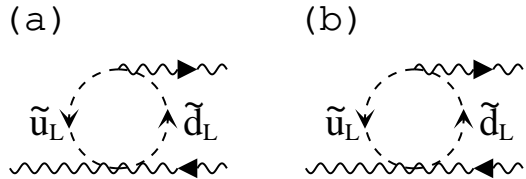,angle=0,height=2.5cm,silent=0}
\end{center}
\caption{Graphs with scalar-fermion seagull (SFSG) vertices that contribute 
to $WW\gamma$ and $WWZ$ vertex corrections.  Because of the coupling to a $W$ 
boson, only the left-handed scalar fermions contribute.  The arrows on the 
$W$-boson lines indicate the flow of negative electric charge.}
\label{fig-wwv-sg}
\end{figure}
They too contribute to the $f_i^{\gamma\,(1)}$ and $f_i^{Z\,(1)}$
form factors of Eqn.~(\ref{big-F}), and we parameterize their contributions
by $f_i^{\gamma\,(1)\,{\rm SFSG}}$ and $f_i^{Z\,(1)\,{\rm SFSG}}$.  The
explicit expressions are contained in Appendix~\ref{app-vertices}.
Since the two-point-function corrections are treated separately we write
\begin{eqnarray}
f_i^{V\,(1)} & = & 
  f_i^{V\,(1)\,{\rm SFT}} + f_i^{V\,(1)\,{\rm SFSG}} \;,
\end{eqnarray}
for $V = \gamma,\,Z$.  There are neither boxes nor $t$-channel vertex 
corrections that contain only scalar fermions but no other non-SM 
particles in the loop.

Next we turn to the calculation of the amplitudes where one $W$ boson is 
replaced by a Goldstone boson.  The propagator corrections were already 
discussed above.
The triangle-type graphs are shown in 
Fig.~\ref{fig-vwx-triangle}.
\begin{figure}[tbhp]
\begin{center}
\leavevmode\psfig{file=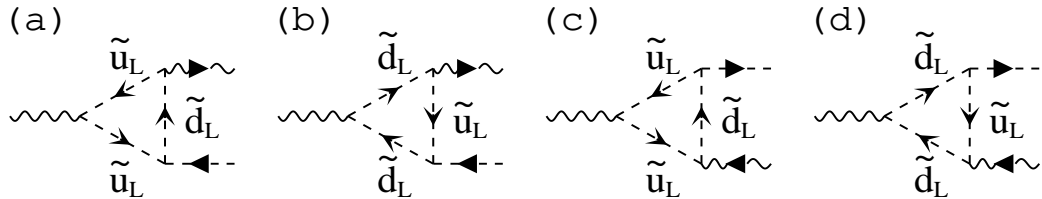,angle=0,width=14cm,silent=0}
\end{center}
\caption{Scalar-fermion triangle-type (SFT) Feynman diagrams. Diagrams (a) and 
(b) contribute to $W^-\chi^+$ production while (c) and (d) contribute to 
$W^+\chi^-$ production.  The arrows on the $W$-boson and Goldstone-boson lines 
indicate the flow of negative electric charge.}
\label{fig-vwx-triangle}
\end{figure}
The Feynman diagrams in Fig.~\ref{fig-vwx-triangle}(a) and (b) contribute to 
the $h_i^{\gamma\,(1)}$ and $h_i^{Z\,(1)}$ form factors of Eqn.~(\ref{big-H})
while those in Fig.~\ref{fig-vwx-triangle}(c) and (d) contribute to 
$\overline{h}_i^{\gamma\,(1)}$ and $\overline{h}_i^{Z\,(1)}$.  To distinguish 
these contributions from those of the other graphs we use the notation
$h_i^{\gamma\,(1)\,{\rm SFTcd tes}}$, $h_i^{Z\,(1)\,{\rm SFT}}$, 
$\overline{h}_i^{\gamma\,(1)\,{\rm SFT}}$ and 
$\overline{h}_i^{Z\,(1)\,{\rm SFT}}$.  There are also seagull-type diagrams 
as shown in Fig.~\ref{fig-vwx-sg}.
\begin{figure}[tbhp]
\begin{center}
\leavevmode\psfig{file=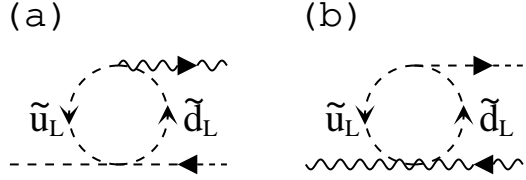,angle=0,height=2.5cm,silent=0}
\end{center}
\caption{Scalar-fermion seagull-type (SFSG) Feynman diagrams.  Diagram (a) 
contributes to $W^-\chi^+$ production while diagram (b) contributes to 
$W^+\chi^-$ production.  The arrows on the $W$-boson and Goldstone-boson lines 
indicate the flow of negative electric charge.}
\label{fig-vwx-sg}
\end{figure}
Fig.~\ref{fig-vwx-sg}(a) contributes to 
$h_i^{\gamma\,(1)\,{\rm SFSG}}$ and $h_i^{Z\,(1)\,{\rm SFSG}}$ while  
Fig.~\ref{fig-vwx-sg}(b) contributes to 
$\overline{h}_i^{\gamma\,(1)\,{\rm SFSG}}$ and 
$\overline{h}_i^{Z\,(1)\,{\rm SFSG}}$.  Then,
\begin{subequations}
\begin{eqnarray}
h_i^{V\,(1)} & = & h_i^{V\,(1)\,{\rm SFT}} 
+ h_i^{V\,(1)\,{\rm SFSG}} \;,\\
\overline{h}_i^{V\,(1)} & = & \overline{h}_i^{V\,(1)\,{\rm SFT}}
+ \overline{h}_i^{V\,(1)\,{\rm SFSG}}\;.
\end{eqnarray}
\end{subequations}
There are neither boxes nor $t$-channel vertex-correction diagrams which
need to be considered.

We are now ready to study the BRS sum rules among the above loop contributions.
For simplicity we focus on the two sum rules given by 
$\overline{G}_{1,\tau}= 0$.  The contributions from the $WW\gamma$ and 
$W\chi\gamma$ vertices enter as 
\begin{eqnarray}
\nonumber
\overline{G}_{1,\tau}^\gvtx & = & \sum_i
\overline{\xi}_{i1}F_{i,\tau}^\gvtx
     - C_{\rm mod}\, \overline{H}_{1,\tau}^\gvtx\\
\nonumber
& = & \frac{Q_e\hatesq}{s} \Big( \sum_i \overline{\xi}_{i1} f_i^{\gamma\,(1)}
     - \overline{h}_1^{\gamma\,(1)} \Big)\\
& = & \frac{Q_e\hatesq}{s\mwsq}\bigg\{-\pitww(\mwsq) + \pitgg(s)
+ \frac{\hatc}{\hats} \pitgz(s) \bigg\}\;,
\label{last-line-g}
\end{eqnarray}
while the contributions of the $WWZ$ and $W\chi Z$ vertices are given by
\begin{eqnarray}
\nonumber
\overline{G}_{1,\tau}^\zvtx & = & \sum_i \overline{\xi}_{i1}F_{i,\tau}^\zvtx
     - C_{\rm mod}\,\overline{H}_{1,\tau}^\zvtx\\
\nonumber
& = & \frac{(T^3_e-\hatssq Q_e)\hatgsq}{s-\mwsq/\hat{c}^2} 
    \Big( \sum_i \overline{\xi}_{i1} f_i^{Z\,(1)}
     - \frac{\hatssq}{\hatcsq}\overline{h}_1^{Z\,(1)} \Big)\\
& = &\frac{(T^3_e-\hatssq Q_e)\hatgsq}{(s-\mwsq/\hat{c}^2)\mwsq}
\bigg\{ -\pitww(\mwsq) + \pitzz(s) + \frac{\hats}{\hatc}\pitgz(s)\bigg\}  \;. 
\label{last-line-z}
\end{eqnarray}
The last line in Eqn.~(\ref{last-line-g}) and the last line in 
Eqn.~(\ref{last-line-z}) are obtained by explicit calculation; see 
Appendix~\ref{app-details}.  At this point we observe that we have reproduced 
Eqns.~(\ref{ggVTX}) and (\ref{gzVTX}).  Hence, we can straightforwardly adopt 
the results of Section~\ref{sec-brs-loop} and conclude
\begin{equation}
\overline{G}_{1,\tau}^\born + \overline{G}_{1,\tau}^\nc + 
\overline{G}_{1,\tau}^\gvtx +  \overline{G}_{1,\tau}^\zvtx = 0\;.
\end{equation}
The cancellation of many three-point functions against a few two-point 
functions in this weighted sum provides an excellent test of our analytic 
and numerical calculations.  In general the relation $\overline{G}_{1,\tau}= 0$
tests six of the form factors that contribute to physical processes.
They are $F_{3,\tau}$, $F_{4,\tau}$ and $F_{8,\tau}$ for both electron 
helicities; see Eqn.~(\ref{eq-expbrs2a}).  In this particular calculation
$F_{4,\tau} = 0$ by CP invariance, and there is no one-loop contribution to 
$F_{8,\tau}$ where the scalar fermions are the only non-SM particles to 
contribute.

For $\overline{G}_{2,\tau}$ the Born and NC contributions drop out trivially.  
The only nontrivial piece is a relation between the $f_i^{V\,(1)}$ and 
$\overline{h}_i^{V\,(1)}$ given by
\begin{subequations}
\begin{eqnarray}
\sum_i 
\overline{\xi}_{i2} f_i^{\gamma\,(1)} - \overline{h}_2^{\gamma\,(1)} &=& 0\;,\\
\sum_i 
\overline{\xi}_{i2} f_i^{Z\,(1)}
     + \frac{\hatssq}{\hatcsq}\overline{h}_2^{Z\,(1)} & = & 0 \;,
\end{eqnarray}
\end{subequations}
which reduces to the following identity among three-point functions:
\begin{equation}
\sum_i C_i^{\rm SF}(p_1,p_2,m_1^2,m_2^2,m_1^2) \overline{\xi}_{i2}
- \frac{m_1^2-m_2^2}{m_W}\overline{c}_2^{\rm SF}(p_1,p_2,m_1^2,m_2^2,m_1^2)
= 0\;,
\end{equation}
where the $C_i^{\rm SF}$ and $\overline{c}_i^{\rm SF}$ are linear combinations
of three-point integral functions given in Appendix~\ref{app-vertices}.
In general the sum rules $\overline{G}_{2,\tau}= 0$ test four physical form 
factors, $F_{1,\tau}$, $F_{2,\tau}$, $F_{3,\tau}$ and $F_{4,\tau}$; see 
Eqn.~(\ref{eq-expbrs2b}).  In this example $F_{4,\tau}=0$ by CP invariance.

Up to the overall normalization $\overline{G}_{3,\tau}$ reduces to the 
tree-level result.  This is because $F_{5,\tau}$ and $F_{6,\tau}$ do not 
receive contributions from the scalar-fermion loops, and there is no one-loop
contribution to $F_{8,\tau}$ and $F_{9,\tau}$ where scalar fermions are the 
only non-SM particles to contribute.  This concludes our analytical 
demonstration for all six $\overline{G}_{i,\tau} = 0$ sum rules.  
Employing the scheme outlined at the beginning of Section~\ref{sec-ff} 
we have successfully verified the sum rules numerically to sixteen decimal 
places with a Fortran program written in double precision.
The remaining sum rules,
$G_{i,\tau} = 0$, are verified similarly.  However, since all of the physical
form factors that receive one-loop scalar-fermion corrections (that are not 
purely a change in the overall normalization) have already been tested by 
$\overline{G}_{1,\tau} = 0$ and $\overline{G}_{2,\tau} = 0$, there is no 
need to additionally verify that $G_{i,\tau} = 0$.


\section{Conclusions}
\label{sec-conclusions}
\cleqn

In this paper we have systematically studied sum rules which follow 
from BRS invariance and can be used to test form factors for the 
process $e^+e^- \rightarrow W^+W^-$.  The BRS invariance of the quantum
gauge field theory leads to identities among amplitudes involving
$W$ bosons and amplitudes where some of those $W$ bosons have been 
replaced by their associated Nambu-Goldstone bosons.  We have employed
the most general form-factor decompositions of the $e^+e^- \rightarrow 
W^+W^-$, $e^+e^- \rightarrow \chi^\pm W^\mp$ and 
$e^+e^- \rightarrow \chi^+ \chi^-$ amplitudes to rewrite the identities 
among matrix elements as sum rules among form factors.  The sum rules
may be applied order by order in perturbation theory, and hence they may
be used to verify the correctness of perturbative calculations.  We have
provided a thorough discussion at the tree and one-loop levels.  There 
are eighteen form factors which contribute to physical $e^+e^- \rightarrow 
W^+W^-$ amplitudes, but additional form factors for unphysical 
$e^+e^- \rightarrow W^+W^-$ amplitudes and $e^+e^- \rightarrow \chi^\pm W^\mp$
amplitudes must also be calculated in order to apply the `single' BRS sum 
rules.  We conclude that sixteen of the eighteen physical form factors may be 
tested by our sum rules. Two CP-odd form factors, $F_{7,+}$ and $F_{7,-}$, 
decouple from all of the sum rules.

We have explored the relationship between the BRS identities and the 
Goldstone-boson equivalence theorem.  While the equivalence theorem may 
be used to check calculations it is valid only in the high-energy limit.
On the other hand, the BRS sum rules are {\em exact} at all energies.  It 
is hence clear that, of the two techniques, the BRS sum rules provide a far 
more powerful tool for testing perturbative calculations.  We also note that,
because they test directly the full scattering amplitudes that contribute to 
cross sections, the BRS sum rules are more convenient for testing both 
analytic and numerical results than are the Ward-Takahashi identities among 
Green's functions.

Initially we apply the sum rules to the complete set of one-loop contributions.
Aided by the Ward-Takahashi identities 
we demonstrate that the full set of diagrams can 
be subdivided into smaller gauge-invariant subsets, and the sum rules may be 
separately applied to each subset.  One such subset is comprised of diagrams
which include a two-point-function correction for an external 
scalar-polarized massive gauge boson or a Goldstone boson; for the calculation 
of physical form factors these effects may be neglected.  The 
wave-function renormalization factors for the final-state physical $W$ bosons
also drop out from our sum sum rules as a part of the overall normalization
factor.  A third subset includes propagator corrections for the neutral 
gauge bosons, vertex corrections for the $VW^+W^-$ and $V\chi^\pm W^\mp$ 
vertices ($V = \gamma, Z$) and the $W$- and $Z$-boson mass shifts.
The diagrams that remain are 
corrections to the $eeV$ vertices, the $ee$ two-point-function corrections,  
corrections to the $e\nu W$ vertex and the neutrino propagator 
as well as box corrections; these diagrams form a fourth subset.

We have argued that our sum rules are useful for testing higher-order
perturbative calculations.  As an illustration we have discussed the 
contributions of a doublet of MSSM scalar fermions to the one-loop form 
factors for $e^+e^- \rightarrow W^+W^-$ helicity amplitudes.  
For this example we have demonstrated analytically that the sum rules 
are satisfied. We note that we have also been successful in verifying the 
sum rules numerically.  The sum rules
will be used extensively in our ongoing work\cite{achksu} where we complete 
the one-loop calculation\cite{alam94} of $e^+e^- \rightarrow W^+W^-$ helicity 
amplitudes in the MSSM and perform the related phenomenological studies. 
In Ref.~\cite{achksu} we demonstrate how the BRS sum rules 
may be satisfied to the limit of floating-point precision.

\section*{Acknowledgments}

The authors would like to thank G.C.~Cho and S.~Ishihara 
for useful discussions.  We would 
also like to thank Kenichi Hikasa for sharing his personal notes on the 
Lagrangian for the supersymmetric Standard Model.  
This work is supported, in 
part, by Grant-in-Aid for Scientific Research from the Ministry of Education, 
Science and Culture of Japan.  The work of R.~Szalapski is also supported in 
part by the National Science Foundation (NSF) through grant number INT9600243.

\newpage
\appendix


\section{Gauge-boson two-point functions}
\label{app-propagators}
\cleqn

Here we present the scalar-fermion contributions to the gauge-boson 
propagator at one loop.  In this paper we restrict the discussion to a 
single SU(2)$_L$ doublet of scalar fermions, $(\tilde{u}_L, \tilde{d}_L)^T$, 
and their right-handed singlet counterparts, $\tilde{u}_R$, $\tilde{d}_R$,
with no left-right mixing.  Complete and general results for the MSSM will 
be presented in Ref.~\cite{achksu}.  The propagator corrections are 
renormalized in the $\overline{\rm MS}$ scheme.  They can be decomposed 
as\cite{hhkm94} 
\begin{eqnarray}
  \Pi_{\mu\nu}^{AB} (q^2)
  = \left( - g^{\mu\nu} +  \frac{q_\mu q_\nu}{q^2} \right) \Pi_T^{AB} (q^2)
    + \frac{q_\mu q_\nu}{q^2}\, \Pi_L^{AB} (q^2)\;.
\end{eqnarray}
Up to the contributions of tadpole diagrams we find the one-loop scalar-fermion
contributions to the transverse gauge-boson propagators are given by
\begin{subequations}
\begin{eqnarray}
  \label{pitgg-sf}
  \Delta \pitgg(\qsq) &=& 
   N_c \frac{\hatesq}{16\pi^2} 
    \sum_{\scferl,\scferr} Q_{\scfer}^2 B_5(q^2,m^2_{\scfer},m^2_{\scfer})\;,\\
  \label{pitgz-sf}
  \Delta \pitgz(\qsq) &=& 
    N_c \frac{\hate\hatgz}{16\pi^2}
    \sum_{\scferl,\scferr} Q_\scfer (T^3_\scfer - \hatssq Q_\scfer)
           B_5(\qsq,m^2_\scfer,m^2_\scfer)\;, \\
  \label{pitzz-sf}
  \Delta \pitzz(\qsq) &=& 
    N_c \frac{\hatgzsq}{16\pi^2}
    \sum_{\scferl,\scferr} (T^3_\scfer - \hatssq Q_\scfer)^2
           B_5(\qsq,m^2_\scfer,m^2_\scfer)\;, \\
  \label{pitww-sf}
  \Delta \pitww(\qsq) &=& \frac{N_c}{2}
    \frac{\hatgsq}{16\pi^2} 
              B_5(\qsq,m^2_\scupl,m^2_\scdownl)\;,
\end{eqnarray}
\end{subequations}
where $\scferl = \scupl, \scdownl$, $\scferr = \scupr, \scdownr$ 
and\cite{hhkm94}
\begin{eqnarray}
  B_5(\qsq,m^2_1,m^2_2) = A(m_1^2) + A(m_2^2) - 4 B_{22}(\qsq,m_1^2,m_2^2)\;.  
  \label{hhkm-b5}
\end{eqnarray}
It is sometimes convenient to express the transverse components of the 
$\Pi$-functions as
\begin{subequations}
\begin{eqnarray}
  \label{pitgg}
   \pitgg(\qsq) &=& \hatesq  \Pi_T^{QQ}(\qsq)\;, \\
  \label{pitgz}
  \pitgz(\qsq)  &=& \hate \hatgz 
     \left\{ \Pi_T^{3Q}(\qsq) - \hatssq \Pi_T^{QQ}(\qsq) \right\}\;, \\
  \label{pitzz}
  \pitzz(\qsq) &=& \hatgzsq 
     \left\{ \Pi_T^{33}(\qsq) - 2 \hatssq \Pi_T^{3Q}(\qsq) 
             + \hat{s}^4 \Pi_T^{QQ}(\qsq) \right\}\;, \\
  \label{pitww}
  \pitww(\qsq) &=& \hatgsq  \Pi_T^{11}(\qsq)\;.
\end{eqnarray}
\end{subequations}
The propagator functions $\Pi^{QQ}$, $\Pi^{3Q}$, $\Pi^{33}$ and $\Pi^{11}$
are then free of the coupling factors that appear in 
Eqns.~(\ref{pitgg-sf})-(\ref{pitww-sf}).


\section{Ward identities among the unphysical two-point functions}
\label{app-unphysical}
\cleqn

Here we would like to show Eqn.~(\ref{unphys2}).
First, we express the tree-level amplitudes for 
$e^+e^- \rightarrow W^+_PW^-_S$ as 
\begin{eqnarray}
  {\cal M}(e^+e^- \rightarrow W^+_PW^-_S)
    =  {\cal M}(\gamma) + {\cal M}(Z) + {\cal M}(t)\;,
\end{eqnarray}
where ${\cal M}(V)$ represents a diagram with the $s$-channel exchange of
a $V$-boson where $V = \gamma, Z$, and ${\cal M}(t)$ denotes the $t$-channel
neutrino-exchange diagram. (See Fig.~\ref{fig-eeww-tree}.)  
Second, the amplitude for 
$e^+e^- \rightarrow W^+_P\chi^-$ can be expressed as
\begin{eqnarray}
  {\cal M}(e^+e^- \rightarrow W^+_P\chi^-)
    =  \widetilde{{\cal M}}(\gamma) + \widetilde{{\cal M}}(Z)\;,
\end{eqnarray}
with $\widetilde{{\cal M}}(V)$ representing the diagram with the $s$-channel 
exchange of a $V$-boson for $V = \gamma, Z$. (See Fig.~\ref{fig-eewx-tree}.)  
Then the BRS identity of 
Eqn.~(\ref{eq-brs2}) implies
\begin{eqnarray}
\nonumber
{\cal M}(e^+e^- \rightarrow W^+_PW^-_S) 
+ i {\cal M}(e^+e^- \rightarrow W^+_P \chi^-) \makebox[-4cm]{} &&\\
& = & {\cal M}(\gamma) + {\cal M}(Z) + {\cal M}(t) 
   + i \left\{ \widetilde{{\cal M}}(\gamma) + \widetilde{{\cal M}}(Z)\right\} 
   = 0\;,
\label{treeid}
\end{eqnarray}
at the tree level.

At the one-loop level the $W_S W_S$ and $\chi W_S$ propagator corrections 
contribute to the $F_{i,\tau}^{\unphys}$ form factors for 
$e^+e^-\rightarrow W^+_P W^-_S$ amplitudes through the Feynman diagrams of 
Fig.~\ref{fig-chi-ws}.  
\begin{figure}[tbhp]
\begin{center}
\leavevmode\psfig{file=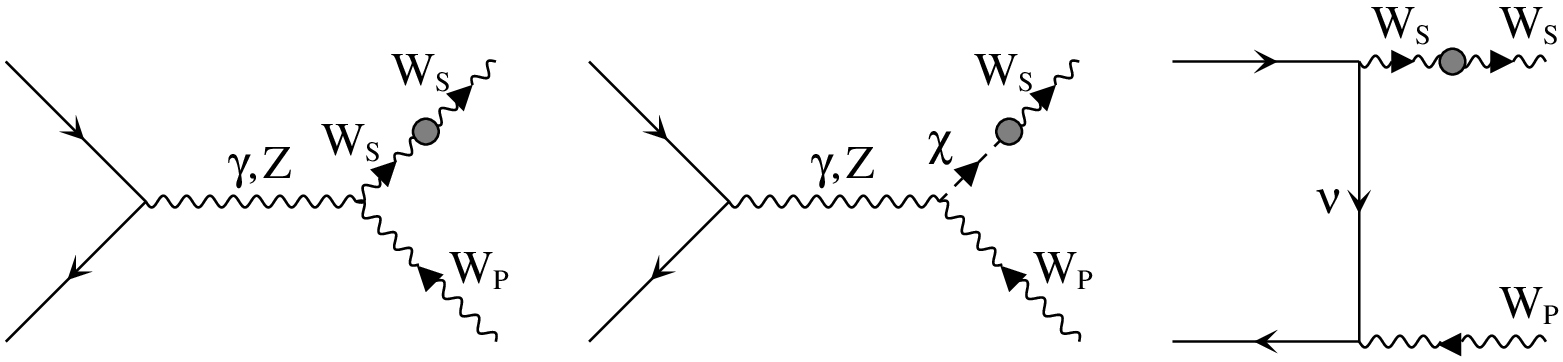,angle=0,height=4cm,silent=0}
\end{center}
\caption{Contributions to $e^+e^-\rightarrow W^+_P W^-_S$ amplitudes from 
corrections to the external lines for unphysical scalars.  The arrows on 
the $W$- and Goldstone-boson lines indicate the flow of negative electric 
charge.}
\label{fig-chi-ws}
\end{figure}
We find
\begin{eqnarray}
&&  {\cal M}^\unphys(e^+e^- \rightarrow W^+_PW^-_S) \nonumber\\
&& = \left\{ {\cal M}(\gamma) + {\cal M}(Z) + {\cal M}(t) \right\} 
     \Pi_{S,W_S}^{W_S W_S}(m_W^2) +  
     \left\{ \widetilde{{\cal M}}(\gamma) + \widetilde{{\cal M}}(Z) \right\} 
     \Pi_{S,W_S}^{\chi W_S}(m_W^2)\;,
\end{eqnarray}
where $\Pi_{S,W_S}^{\phi_1 \phi_2}(p^2)$ 
is the propagator function for the unphysical particles 
$\phi_1$ and $\phi_2$ ($\phi_i =  W_S$, $\chi$) which is defined by  
\begin{eqnarray}
   \Pi_{S,W_S}^{\phi_1, \phi_2} (p^2) 
  \equiv \frac{\Pi^{\phi_1, \phi_2} (p^2) - 
                \Pi^{\phi_1, \phi_2} (\xi m_W^2) }
              {p^2 - \xi m_W^2}\;, 
\end{eqnarray} 
where $\xi m_W^2$ is the common mass-squired of the $\phi_i$ scalars.
Fig.~\ref{fig-ws-chi} shows the Feynman diagrams that include corrections 
to the external lines for unphysical scalars and contribute to the 
$H_{j,\tau}^{\unphys}$ form factors for the 
$e^+e^-\rightarrow W^+_P \chi^-$ amplitudes.  
\begin{figure}[tbhp]
\begin{center}
\leavevmode\psfig{file=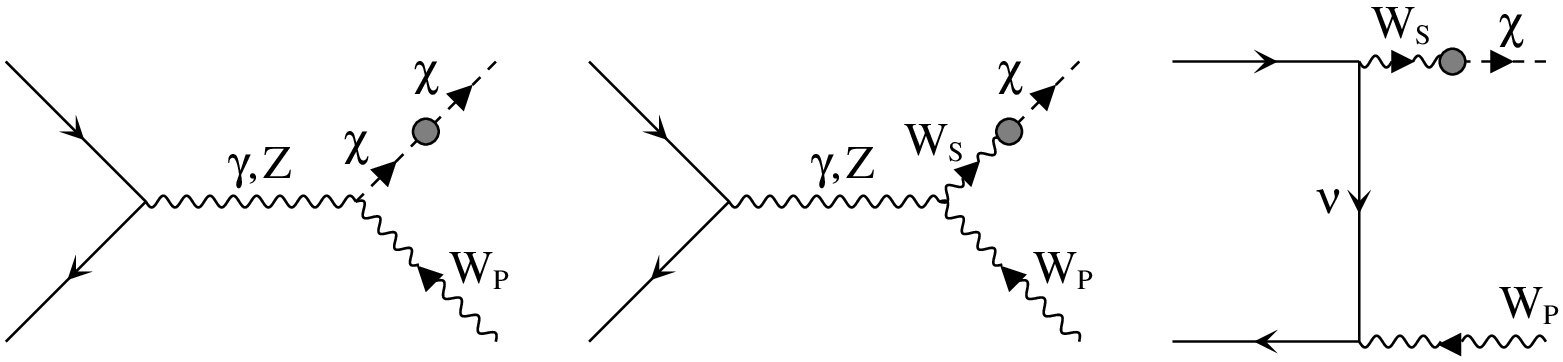,angle=0,height=4cm,silent=0}
\end{center}
\caption{Contributions to $e^+e^-\rightarrow W^+_P \chi^-$ amplitudes from 
corrections to the external lines for unphysical scalars.  The arrows on 
the $W$- and Goldstone-boson lines indicate the flow of negative electric 
charge.}
\label{fig-ws-chi}
\end{figure}
We find
\begin{eqnarray}
&&  {\cal M}^\unphys(e^+e^- \rightarrow W^+_P \chi^-) \nonumber \\
&& =   \left\{ \widetilde{{\cal M}}(\gamma) + \widetilde{{\cal M}}(Z) \right\} 
     \Pi_{S,W_S}^{\chi \chi}(m_W^2)  + 
\left\{ {\cal M}(\gamma) + {\cal M}(Z) + {\cal M}(t) \right\} 
     \Pi_{S,W_S}^{W_S \chi}(m_W^2) \;.  
\end{eqnarray}
Therefore,
\begin{eqnarray}
&&     {\cal M}^\unphys(e^+e^- \rightarrow W^+_PW^-_S)
 + i {\cal M}^\unphys(e^+e^- \rightarrow W^+_P\chi^-) \nonumber \\
&&=   \left\{ {\cal M}(\gamma) + {\cal M}(Z) + {\cal M}(t) \right\} 
      \left\{ \Pi_{S,W_S}^{W_SW_S}(m_W^2)   
      + i \Pi_{S,W_S}^{W_S \chi}(m_W^2)  \right\} \nonumber\\
&& \;\;\;\;\;\;\;\;\;\;\;\;\;\;\;\;
    + \left\{  \widetilde{{\cal M}}(\gamma) 
      + \widetilde{{\cal M}}(Z) \right\} 
      \left\{  \Pi_{S,W_S}^{\chi W_S}(m_W^2)  
      + i \Pi_{S,W_S}^{\chi \chi}(m_W^2) \right \} 
\nonumber \\
&&=  \left\{ {\cal M}(\gamma) + {\cal M}(Z) + {\cal M}(t) \right\} 
  \left\{  \Pi_{S,W_S}^{W_SW_S}(m_W^2)  + i \Pi_{S,W_S}^{W_S \chi}(m_W^2) 
          + i \Pi_{S,W_S}^{\chi W_S}(m_W^2)  
          -  \Pi_{S,W_S}^{\chi \chi}(m_W^2) \right\}, \nonumber \\ 
\label{upid}
\end{eqnarray}
where we have used the tree-level identity in Eqn.~(\ref{treeid}). 
The right-hand side of Eqn.~(\ref{upid}) is identically zero by the following 
Ward identity for the unphysical two-point functions\cite{lv80}:
\begin{eqnarray}
  \Pi^{W_SW_S}(\qsq) + i \,\Pi^{W_S\chi}(\qsq) + 
   i \,\Pi^{\chi W_S}(\qsq) - 
    \Pi^{\chi\chi}(\qsq) = 0\;.
\end{eqnarray}
Finally, by expanding the left-hand side of Eqn.~(\ref{upid}) in the 
$\overline{S}^{\mu\beta}_j$ tensors we obtain Eqn.~(\ref{unphys2}).


\section{One-loop vertex corrections}
\label{app-vertices}
\cleqn

The propagator corrections from scalar-fermion loops were presented in 
Appendix~\ref{app-propagators}.  In this section we will present the vertex 
corrections from the scalar fermions $(\tilde{u}_L, \tilde{d}_L)^T$, 
$\tilde{u}_R$ and $\tilde{d}_R$.  Complete and general results for the MSSM 
will be presented in Ref.~\cite{achksu}.  

\subsection{$WW\gamma$ and $WWZ$ vertex corrections}

We begin by assigning masses and momenta as in Fig.~\ref{fig-wwv-triangle-p}.
\begin{figure}[tbhp]
\begin{center}
\leavevmode\psfig{file=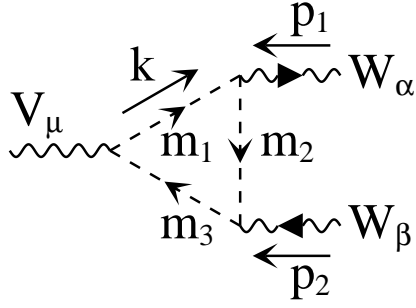,angle=0,height=4cm,silent=0}
\end{center}
\caption{Mass and momentum assignments for the calculation of the 
scalar-fermion triangle-type (SFT) loop corrections to $WW\gamma$ and $WWZ$ 
vertices.  All momenta are incoming.  The arrows along the $W$-boson lines 
indicate the flow of negative electric charge.}
\label{fig-wwv-triangle-p}
\end{figure}
Note that in the definition of the tensors $T_i^{\mu\alpha\beta}$ of 
Eqns.~(\ref{ff_T1})-(\ref{ff_T16}) we used the outgoing $W$-boson momenta
$p$ and $\overline{p}$.  See Fig.~\ref{fig-eeww-blob}.  In order to be 
consistent with the notation of the Fortran package FF\cite{gjo91} and that 
of Ref.~\cite{hhkm94} we evaluate our loop integrals using the incoming
momenta $p_1 = -p$ and $p_2 = -\overline{p}$.

If we neglect the coupling factors and employ arbitrary masses $m_1$, $m_2$
and $m_3$, then the calculation of the graph in Fig.~\ref{fig-wwv-triangle-p},
yields the following tensor structure
\begin{equation}
T_{\rm SFT}^{\mu\alpha\beta}(p_1,p_2,m_1^2,m_2^2,m_3^2) = 
\sum_i C^{\rm SF}_i(p_1,p_2,m_1^2,m_2^2,m_3^2) T_i^{\mu\alpha\beta}\;,
\end{equation}
where the nonzero $C^{\rm SF}_i$ are given by
\begin{subequations}
\begin{eqnarray}
\label{C1sft}
C^{\rm SF}_1(p_1,p_2,m_1^2,m_2^2,m_3^2) & = & 
  4 (C_{36}-C_{35})\;, \\
\label{C2sft}
C^{\rm SF}_2(p_1,p_2,m_1^2,m_2^2,m_3^2) & = &
  4\mwsq (C_{33} - C_{34} + C_{23} - C_{22})\;,\\
\label{C3sft}
C^{\rm SF}_3(p_1,p_2,m_1^2,m_2^2,m_3^2) & = & 
  4  (C_{35}-C_{36}+C_{24})\;,\\
\label{C10sft}
C^{\rm SF}_{10}(p_1,p_2,m_1^2,m_2^2,m_3^2) & = &  
  4  (2C_{36}-2C_{35}-C_{24})\;,\\
\nonumber
C^{\rm SF}_{11}(p_1,p_2,m_1^2,m_2^2,m_3^2) & = &  
  2\mwsq (4C_{33}-2C_{34}-2C_{31}+5C_{23} \\
\label{C11sft}
&&\makebox[2cm]{} -2C_{22}-3C_{21} +C_{12}-C_{11})\;,\\
\label{C13sft}
C^{\rm SF}_{13}(p_1,p_2,m_1^2,m_2^2,m_3^2) & = & -C^{\rm SF}_{10} \;,\\
\label{C14sft}
C^{\rm SF}_{14}(p_1,p_2,m_1^2,m_2^2,m_3^2) & = &  
  2\mwsq (4C_{34}-2C_{33}-2C_{32}+C_{22}-C_{23})\;,\\
\nonumber
C^{\rm SF}_{16}(p_1,p_2,m_1^2,m_2^2,m_3^2) & = &  
  \mwsq (-12C_{34}+12C_{33}+4C_{32}-4C_{31} \\ 
\label{C16sft}
&&\makebox[2cm]{} +16C_{23}-8C_{22}-8C_{21}+3C_{12}-3C_{11})\;.
\end{eqnarray}
\end{subequations}
Only the the nonvanishing $C_i^{\rm SF}$-functions are listed.
The various $C$-functions on the right-hand side, 
which are discussed in
Appendix~\ref{app-cmunurho}, are assigned the same arguments as the functions 
on the left-hand side.  
The next step is to provide the correct couplings and 
masses and then sum over all triangle graphs.  For the $WW\gamma$ vertex we 
find,
\begin{equation}
f_i^{\gamma\,(1)\,{\rm SFT}} = \frac{N_c}{2} \frac{\hatgsq}{16\pi^2}
\bigg[Q_\scup C_i^{\rm SF}(p_1,p_2,m^2_\scupl,m^2_\scdownl,m^2_\scupl)  
 -Q_\scdown C_i^{\rm SF}(p_1,p_2,m^2_\scdownl,m^2_\scupl,m^2_\scdownl)\bigg]\;,
\label{fg1sft}
\end{equation}
while the result for the $WWZ$ vertex is 
\begin{eqnarray}
\nonumber 
f_i^{Z\,(1)\,{\rm SFT}} & = & \frac{N_c}{2} \frac{\hatgzsq}{16\pi^2}
\bigg[(T^3_\scupl-\hatssq Q_\scup) 
C_i^{\rm SF}(p_1,p_2,m^2_\scupl,m^2_\scdownl,m^2_\scupl) \\ 
&&\makebox[2cm]{}
 -(T^3_\scdownl -\hatssq Q_\scdown)
C_i^{\rm SF}(p_1,p_2,m^2_\scdownl,m^2_\scupl,m^2_\scdownl)\bigg]\;.
\label{fz1sft}
\end{eqnarray}
The superscript `SFT' indicates the scalar-fermion triangle-type correction.
The $f_i^{\gamma\,(1)\,{\rm SFT}}$ form factors of Eqn.~(\ref{fg1sft}) and the 
$f_i^{Z\,(1)\,{\rm SFT}}$ form factors of Eqn.~(\ref{fz1sft})
are nonvanishing whenever the corresponding $C_i^{\rm SF}$ functions of 
Eqns.~(\ref{C1sft})-(\ref{C16sft}) are nonvanishing.

The other type of vertex loop correction includes one scalar-fermion 
seagull-type (SFSG) vertex.  See Fig.~\ref{fig-wwv-sg-p}.
\begin{figure}[tbhp]
\begin{center}
\leavevmode\psfig{file=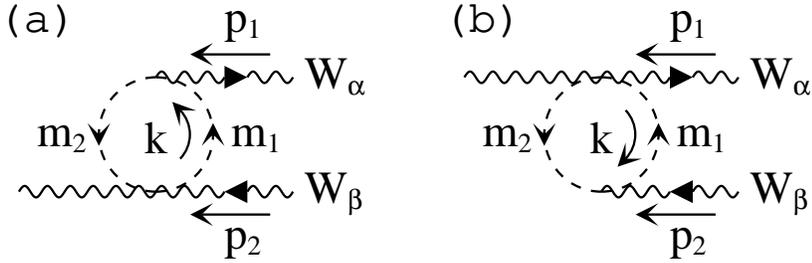,angle=0,height=3.5cm,silent=0}
\end{center}
\caption{Corrections to $WW\gamma$ and $WWZ$ vertices involving one 
scalar-fermion seagull (SFSG) vertex.  These graphs contribute only for 
unphysical polarizations of the $W$ bosons.  The arrows along the $W$-boson
lines indicate the flow of negative electric charge.}
\label{fig-wwv-sg-p}
\end{figure}
These Feynman diagrams are explicitly proportional to the momentum of the 
$W$ boson attached to the three-point vertex in the loop, and hence they 
contribute only when the same $W$-boson has an unphysical scalar polarization.
An explicit calculation yields
\begin{subequations}
\begin{eqnarray}
\label{fg10sfsg}
f_{10}^{\gamma\,(1)\,{\rm SFSG}} & = & -\frac{N_c}{2}\frac{\hatgsq}{16\pi^2}
(Q_\scup+Q_\scdown)\Big[ 2B_1+B_0 \Big](\mwsq,m^2_\scdownl,m^2_\scdownr)\;,\\
\label{fg13sfsg}
f_{13}^{\gamma\,(1)\,{\rm SFSG}} & = & -f_{10}^{\gamma\,(1)\,{\rm SFSG}}\;,
\end{eqnarray}
\end{subequations}
for the $WW\gamma$ vertex.  For the $WWZ$ vertex
we find
\begin{subequations}
\begin{eqnarray}
\label{fz10sfsg}
f_{10}^{Z\,(1)\,{\rm SFSG}} & = & \frac{N_c}{2}\frac{\hatgzsq\hatssq}{16\pi^2}
(Q_\scup+Q_\scdown)\Big[ 2B_1+B_0 \Big](\mwsq,m^2_\scdownl,m^2_\scdownr)\;,\\
\label{fz13sfsg}
f_{13}^{Z\,(1)\,{\rm SFSG}} & = & -f_{10}^{Z\,(1)\,{\rm SFSG}}\;.
\end{eqnarray}
\end{subequations}
Only the nonvanishing form factors are shown.

\subsection{$W^\pm\chi^\mp\gamma$ and $W^\pm\chi^\mp Z$ vertex corrections}

We begin by assigning masses and momenta as in Fig.~\ref{fig-vwx-triangle-p}.
\begin{figure}[tbhp]
\begin{center}
\leavevmode\psfig{file=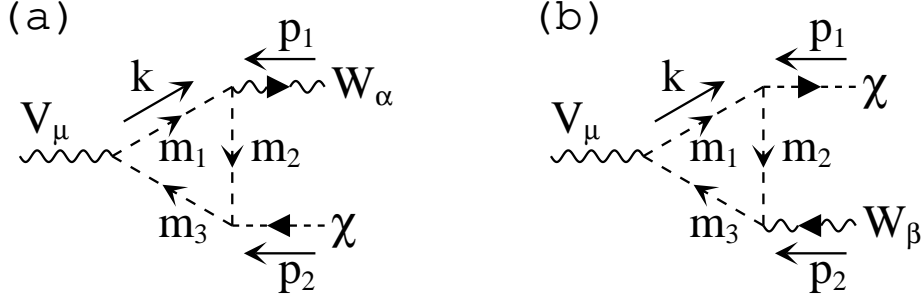,angle=0,height=4cm,silent=0}
\end{center}
\caption{Mass and momentum assignments for the calculation of the 
scalar-fermion triangle-type (SFT) one-loop contributions to 
$W^\pm\chi^\mp\gamma$ and $W^\pm\chi^\mp Z$ vertices.  All momenta are 
incoming.  The arrows along the $W$- and Goldstone-boson lines indicate the 
flow of negative electric charge.}
\label{fig-vwx-triangle-p}
\end{figure}
Neglecting coupling factors and computing the loops yields the following 
tensor structures:
\begin{subequations}
\label{vwx_vxw_tensors}
\begin{eqnarray}
\label{vwx_tensor}
S^{\mu\alpha}_{\rm SFT}(p_1,p_2,m_1^2,m_2^2,m_3^2) & = & 
\sum_i c_i^{\rm SF}(p_1,p_2,m_1^2,m_2^2,m_3^2)S_i^{\mu\alpha}\;,\\
\label{vxw_tensor}
\overline{S}^{\mu\beta}_{\rm SFT}(p_1,p_2,m_1^2,m_2^2,m_3^2) & = & 
\sum_i \overline{c}_i^{\rm SF}(p_1,p_2,m_1^2,m_2^2,m_3^2)
\overline{S}_i^{\mu\beta}\;,
\end{eqnarray}
\end{subequations}
corresponding to Fig.~\ref{fig-vwx-triangle-p}(a) and 
Fig.~\ref{fig-vwx-triangle-p}(b) respectively.  The $c^{\rm SF}$ functions are 
given by
\begin{subequations}
\begin{eqnarray}
c_1^{\rm SF}(p_1,p_2,m_1^2,m_2^2,m_3^2) & = & 4C_{24}/m_W\;,\\
c_2^{\rm SF}(p_1,p_2,m_1^2,m_2^2,m_3^2) & = & 
    2 m_W (C_{22} - C_{23})\;,\\
c_3^{\rm SF}(p_1,p_2,m_1^2,m_2^2,m_3^2) & = & 0\;,\\
c_4^{\rm SF}(p_1,p_2,m_1^2,m_2^2,m_3^2) & = & 
m_W ( 2C_{21}-2C_{22}+C_{11}-C_{12})\;,
\end{eqnarray}
\end{subequations}
and the $\overline{c}^{\rm SF}$ functions are
\begin{subequations}
\begin{eqnarray}
\overline{c}_1^{\rm SF}(p_1,p_2,m_1^2,m_2^2,m_3^2) & = & 4C_{24}/m_W\;,\\
\overline{c}_2^{\rm SF}(p_1,p_2,m_1^2,m_2^2,m_3^2) & = & 
    2 m_W (C_{21} - C_{23} + C_{11} - C_{12})\;,\\
\overline{c}_3^{\rm SF}(p_1,p_2,m_1^2,m_2^2,m_3^2) & = & 0\;,\\
\overline{c}_4^{\rm SF}(p_1,p_2,m_1^2,m_2^2,m_3^2) & = & 
m_W ( 2C_{21}-2C_{22}+3C_{11}-3C_{12})\;.
\end{eqnarray}
\end{subequations}
In both cases the $C$-functions on the right-hand side have the same arguments 
as the functions on the left-hand side.  It can be shown that
\begin{equation}
c_i^{\rm SF}(p_1,p_2,m_1^2,m_2^2,m_3^2) = 
\overline{c}_i^{\rm SF}(p_1,p_2,m_1^2,m_2^2,m_3^2)\;.
\end{equation}
Upon incorporating the correct masses and coupling factors we obtain
\begin{subequations}
\begin{eqnarray}
\nonumber
\overline{h}_i^{\gamma\,(1)\,{\rm SFT}} & = & 
\frac{N_c}{2}\frac{\hatgsq}{16\pi^2} \frac{m^2_\scupl-m^2_\scdownl}{m_W}
 \Big[ Q_\scup 
 \overline{c}_i^{\rm SF}(p_1,p_2,m^2_\scupl,m^2_\scdownl,m^2_\scupl)\\
&&\makebox[5cm]{}
 + Q_\scdown 
 \overline{c}_i^{\rm SF}(p_1,p_2,m^2_\scdownl,m^2_\scupl,m^2_\scdownl)\Big]\;,
\label{barhg1sft}
\\ 
\nonumber
\overline{h}_i^{Z\,(1)\,{\rm SFT}} & = & \frac{N_c}{2}\frac{\hatgzsq}{16\pi^2}
 \frac{m^2_\scupl-m^2_\scdownl}{m_W}
 \Big[ (T^3_\scupl-\hatssq Q_\scup)
  \overline{c}_i^{\rm SF}(p_1,p_2,m^2_\scupl,m^2_\scdownl,m^2_\scupl) \\
&&\makebox[5cm]{}
 + (T^3_\scdownl - \hatssq Q_\scdown)
  \overline{c}_i^{\rm SF}(p_1,p_2,m^2_\scdownl,m^2_\scupl,m^2_\scdownl)\Big]\;,
\label{barhz1sft}
\end{eqnarray}
\end{subequations}
and
\begin{subequations}
\begin{eqnarray}
\label{hg1sft}
h_i^{\gamma\,(1)\,{\rm SFT}} & = & -\overline{h}_i^{\gamma\,(1)\,{\rm SFT}}\;,
\\
\label{hz1sft}
h_i^{Z\,(1)\,{\rm SFT}} & = & -\overline{h}_i^{Z\,(1)\,{\rm SFT}} \;.
\end{eqnarray}
\end{subequations}
The Yukawa coupling introduces a factor of mass-dimension one while the 
$c_i^{\rm SF}$ and $\overline{c}_i^{\rm SF}$ functions introduce one inverse
power of mass.  Hence, the $h_i^{V\,(1)\,{\rm SFT}}$ and
$\overline{h}_i^{V\,(1)\,{\rm SFT}}$ loop functions, $V = \gamma,Z$, are 
dimensionless, as expected.

There are also contributions from the loop diagrams that include a seagull 
vertex.  See Fig.~\ref{fig-vwx-sg-p} for the mass and momentum assignments.
\begin{figure}[tbhp]
\begin{center}
\leavevmode\psfig{file=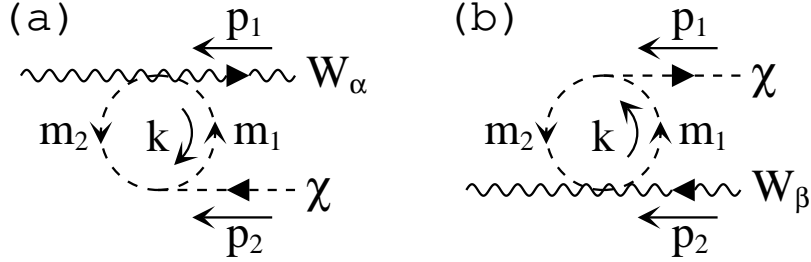,angle=0,height=3.5cm,silent=0}
\end{center}
\caption{One-loop diagrams that contain a scalar-fermion seagull (SFSG) vertex
and contribute to $W^\pm\chi^\mp\gamma$ and $W^\pm\chi^\mp Z$ vertices. The 
arrows along the $W$- and Goldstone-boson lines indicate the flow of negative 
electric charge.}
\label{fig-vwx-sg-p}
\end{figure}
By explicit calculation we find the nonzero contributions are 
\begin{subequations}
\begin{eqnarray}
\label{barh1gsfsg}
\overline{h}_1^{\gamma\,(1)\,{\rm SFSG}} & = & -\frac{N_c}{2}
\frac{\hatgsq}{16\pi^2} \frac{m^2_\scupl-m^2_\scdownl}{\mwsq}
(Q_\scup+Q_\scdown) B_0(\mwsq,m^2_\scdownl,m^2_\scupl)\;,\\
\label{barh1zsfsg}
\overline{h}_1^{Z\,(1)\,{\rm SFSG}} & = & \frac{N_c}{2}
\frac{\hatgzsq \hat{s}^2}{16\pi^2}\frac{m^2_\scupl-m^2_\scdownl}{\mwsq}
(Q_\scup+Q_\scdown) B_0(\mwsq,m^2_\scdownl,m^2_\scupl)\;,
\end{eqnarray}
\end{subequations}
and
\begin{subequations}
\begin{eqnarray}
h_1^{\gamma\,(1)\,{\rm SFSG}} & = & -\overline{h}_1^{\gamma\,(1)\,{\rm SFSG}} 
\;,\\
h_1^{Z\,(1)\,{\rm SFSG}} & = & -\overline{h}_1^{Z\,(1)\,{\rm SFSG}} \;.
\end{eqnarray}
\end{subequations}


\section{BRS sum rules at one loop}
\label{app-details}
\cleqn

The contributions of the vertex corrections to the BRS sum rules involve
complicated algebra.  Many of the details are given in this appendix.  
We begin with some identities involving the three-point tensor integrals.
Starting from Eqn.~(\ref{c3n-reduction}) of Appendix~\ref{app-cmunurho} we 
specialize to the case $m_1 = m_3$, $p_1^2 = p_2^2 = \mwsq$ and $(p_1+p_2)^2 = 
s$. We find
\begin{subequations}
\begin{eqnarray}
\nonumber
\lefteqn{\Big[C_{35} - C_{36} + C_{24}\Big](p_1,p_2,m_1^2,m_2^2,m_1^2) = }&&\\
&&\frac{1}{s-4\mwsq}\bigg\{
2\Big[B_{22}^{(12)}-B_{22}^{(13)}\Big]+2\big(m_1^2-m_2^2-\mwsq\big)
C_{24}(p_1,p_2,m_1^2,m_2^2,m_1^2)\bigg\}\;,\\
\nonumber
\lefteqn{\Big[2C_{35}-2C_{36} + C_{24}\Big](p_1,p_2,m_1^2,m_2^2,m_1^2) = }&&\\
&&\frac{1}{s-4\mwsq}\bigg\{
4\Big[B_{22}^{(12)}-B_{22}^{(13)}\Big]+\big(4m_1^2-4m_2^2-s\big)
C_{24}(p_1,p_2,m_1^2,m_2^2,m_1^2)\bigg\}\;,
\end{eqnarray}
\end{subequations}
where the shorthand notation for the $B$ functions is also given in 
Appendix~\ref{app-cmunurho}.  Using these identities and 
Eqns.~(\ref{C1sft})-(\ref{C16sft}) we find
\begin{subequations}
\begin{eqnarray}
\makebox[-1cm]{}
\sum_{i=1}^{16}
\overline{\xi}_{i1}C_i^{\rm SF}(p_1,p_2,m_1^2,m_2^2,m_1^2)
& = & 2\gamma^2 C_3^{\rm SF}(p_1,p_2,m_1^2,m_2^2,m_1^2) 
+ C_{10}^{\rm SF}(p_1,p_2,m_1^2,m_2^2,m_1^2)\\
& = & \frac{4}{\mwsq}\bigg\{B_{22}^{(12)} - B_{22}^{(13)}+\big(m_1^2-m_2^2\big)
C_{24}(p_1,p_2,m_1^2,m_2^2,m_1^2)\bigg\}\;.
\end{eqnarray}
\end{subequations}
Incorporating Eqns.~(\ref{fg1sft}) and (\ref{barhg1sft}) leads to 
\begin{eqnarray}
\nonumber
\makebox[-1cm]{}
\sum_{i=1}^{16}
\overline{\xi}_{i1} f_i^{\gamma\,(1)\,{\rm SFT}} - 
\overline{h}_1^{\gamma\,(1)\,{\rm SFT}} & = &
\frac{N_c}{2} \frac{\hatgsq}{16\pi^2}\frac{4}{\mwsq}\bigg\{
Q_\scup\Big[B_{22}(\mwsq,m^2_\scupl,m^2_\scdownl) 
- B_{22}(s,m^2_\scupl,m^2_\scupl)\Big]\\
&&\makebox[2.5cm]{} - Q_\scdown\Big[B_{22}(\mwsq,m^2_\scdownl,m^2_\scupl)
- B_{22}(s,m^2_\scdownl,m^2_\scdownl)\Big]\bigg\}\;,\label{g1gsft}
\end{eqnarray}
while from Eqns.~(\ref{fz1sft}) and (\ref{barhz1sft}) we obtain
\begin{eqnarray}
\nonumber
\makebox[-1cm]{}
\sum_{i=1}^{16}
\overline{\xi}_{i1} f_i^{Z\,(1)\,{\rm SFT}} - 
\overline{h}_1^{Z\,(1)\,{\rm SFT}} & = &\\ \nonumber
&&\makebox[-2cm]{}\frac{N_c}{2}\frac{\hatgzsq}{16\pi^2}\frac{4}{\mwsq}\bigg\{
\big(T^3_\scupl-\hatssq Q_\scup\big)\Big[B_{22}(\mwsq,m^2_\scupl,m^2_\scdownl) 
- B_{22}(s,m^2_\scupl,m^2_\scupl)\Big]\\
&& \makebox[.2cm]{}-\big(T^3_\scdownl-\hatssq Q_\scdown\big)
\Big[B_{22}(\mwsq,m^2_\scdownl,m^2_\scupl)
- B_{22}(s,m^2_\scdownl,m^2_\scdownl)\Big]\bigg\}\;.\label{g1zsft}
\end{eqnarray}
The $C$ functions have completely canceled, and the result depends 
only on $B$ functions.

Another useful identity among loop-integral functions is
\begin{equation}
\mwsq\Big[2B_1+B_0\Big](\mwsq,m_1^2,m_2^2) = A(m_1^2) - A(m_2^2)
-\big(m_1^2-m_2^2 \big) B_0(\mwsq,m_1^2,m_2^2)\;.
\end{equation}
Using this we can show
\begin{subequations}
\begin{eqnarray}
\label{gg1sfsg}
\sum_{i=1}^{16}\overline{\xi}_{i1}
f_{i}^{\gamma\,(1)\,{\rm SFSG}} - \overline{h}_1^{\gamma\,(1)\,{\rm SFSG}}
& = & \frac{N_c}{2}\frac{\hatgsq}{16\pi^2}\frac{1}{\mwsq}(Q_\scup+Q_\scdown)
\bigg\{A(m^2_\scupl) - A(m^2_\scdownl)\bigg\}\;,\\
\label{gz1sfsg}
\sum_{i=1}^{16} \overline{\xi}_{i1}
f_{i}^{Z\,(1)\,{\rm SFSG}} - \overline{h}_1^{Z\,(1)\,{\rm SFSG}}
& = & \frac{N_c}{2}\frac{\hatgzsq\hatssq}{16\pi^2}\frac{1}{\mwsq}
(Q_\scup+Q_\scdown)\bigg\{
A(m^2_\scdownl)-A(m^2_\scupl)\bigg\}\;,
\end{eqnarray}
\end{subequations}
where $f_{10}^{\gamma\,(1)\,{\rm SFSG}}$ and $f_{10}^{Z\,(1)\,{\rm SFSG}}$ 
are the only $f_{i}^{V\,(1)\,{\rm SFSG}}$ form factors to make nonzero 
contributions.  Combining the two types of graphs we write
\begin{subequations}
\begin{eqnarray}
\nonumber
\overline{G}_{1,\tau}^\gvtx & = & -\frac{N_c}{2} \frac{\hatgsq}{16\pi^2}
\frac{Q_e \hatesq}{s\,\mwsq}\bigg\{B_5(\mwsq,m^2_\scupl,m^2_\scdownl) \\&&
\makebox[1cm]{}-Q_\scup B_5(s,m^2_\scupl,m^2_\scupl)
+Q_\scdown B_5(s,m^2_\scdownl,m^2_\scdownl)\bigg\}\;,\label{g1gvert}\\
\nonumber
\overline{G}_{1,\tau}^\zvtx & = & -\frac{N_c}{2}\frac{\hatgsq}{16\pi^2}
\frac{(T^3_e-\hatssq Q_e) \hatgzsq}{(s-\mwsq/\hat{c}^2)\mwsq}
\bigg\{\hatcsq B_5(\mwsq,m^2_\scupl,m^2_\scdownl) \\&&
\makebox[1cm]{}-(T^3_\scupl-\hatssq Q_\scup) B_5(s,m^2_\scupl,m^2_\scupl)
+(T^3_\scdownl-\hatssq Q_\scdown) 
B_5(s,m^2_\scdownl,m^2_\scdownl)\bigg\}\label{g1zvert}\;,
\end{eqnarray}
\end{subequations}
where the $B_5$ function is defined in Eqn.~(\ref{hhkm-b5}).
The next step is to insert factors of $2T^3_\scupl = 1$ and 
$-2T^3_\scdownl = 1$ in appropriate places, and then we can immediately
recognize the two-point functions in the basis of current eigenstates as
given in Eqn.~(\ref{pitgg})-(\ref{pitww}).  Using the $\Pi_T$ functions of
Appendix~\ref{app-propagators} we find
\begin{subequations}
\begin{eqnarray}
\nonumber
\overline{G}_{1,\tau}^\gvtx & = & \frac{Q_e \hatesq}{s}\frac{\hatgsq}{\mwsq}
\bigg\{ -\Pi^{11}_T(\mwsq) + \Pi^{3Q}_T(s) \bigg\}\\
\label{g1g-sf}
& = & \frac{Q_e\hatesq}{s\mwsq}\bigg\{-\pitww(\mwsq) + \pitgg(s)
+ \frac{\hatc}{\hats} \pitgz(s) \bigg\}\;,\\
\nonumber
\overline{G}_{1,\tau}^\zvtx & = & 
\frac{(T^3_e-\hatssq Q_e)\hatgzsq}{s-\mwsq/\hat{c}^2}
\frac{\hatgsq}{\mwsq}\bigg\{-\hatcsq \Pi^{11}_T(\mwsq) - \hatssq \Pi^{3Q}_T(s)
+ \Pi^{33}_T(s)\bigg\}\\
& = & \frac{(T^3_e-\hatssq Q_e)\hatgsq}{(s-\mwsq/\hat{c}^2)\mwsq}
\bigg\{ -\pitww(\mwsq) + \pitzz(s) + \frac{\hats}{\hatc}\pitgz(s)\bigg\}  \;.
\end{eqnarray}
\end{subequations}
The contributions of all the vertex corrections to the BRS sum rules are now
expressed entirely in terms of the gauge-boson propagator corrections.
  

\section{The decomposition of $C^{\mu\nu\rho}$, a rank-three 
         three-point  tensor integral}
\label{app-cmunurho}
\cleqn

For our loop integrals we follow the notation of \cite{hhkm94, gjo91}.  
While 
the tensor reduction of the three-point integrals ($C$-functions) through 
rank two is presented in Appendix~D of Ref.~\cite{hhkm94}, it is necessary to 
present here the reduction of the rank-three $C$-function.  In order to make 
this appendix reasonably self-contained, we also give a brief review of the 
lower $C$-functions and the two-point integrals ($B$-functions).

Our measure in $D=4-2\epsilon$ dimensions, defined as
\begin{equation}
\label{measure}
\overline{d^Dk} = \Gamma(1-\epsilon)(\pi\mu^2)^\epsilon d^Dk\;,
\end{equation}
is the $\overline{\rm MS}$ regularization\cite{msbar}.  Propagator factors 
are defined as 
\begin{subequations}
\label{N1N2N3}
\begin{eqnarray}
N_1 & = & k^2 - m_1^2 + i\epsilon \;,\\
N_2 & = & (k+p_1)^2 - m_2^2 + i\epsilon \;,\\
N_3 & = & (k+p_1+p_2)^2 - m_3^2 + i\epsilon \;,
\end{eqnarray}
\end{subequations}
and it is convenient to introduce the following short-hand notation:
\begin{subequations}
\begin{eqnarray}
f_1 & = & m_2^2 - m_1^2 -p_1^2\;,\\
f_2 & = & m_3^2 - m_2^2 - (p_1+p_2)^2 + p_1^2\;.
\end{eqnarray}
\end{subequations}
The $B$-functions are given by 
\begin{equation}
\label{B-def}
\{B_0,\,B^{\mu},\,B^{\mu\nu} \}(p_1,m_1,m_2) 
= \int\frac{\overline{d^dk}}{i\pi^2}
\frac{\{1,\,k^\mu,\,k^\mu  k^\nu \}}{N_1N_2}\;,
\end{equation}
with the decompositions
\begin{subequations}
\label{B-scalar-def}
\begin{eqnarray}
B^{\mu}(p_1,m_1,m_2) & = & p_1^\mu B_{1}(p_1,m_1,m_2)\;,\\
B^{\mu\nu}(p_1,m_1,m_2) & = & g^{\mu\nu}B_{22}(p_1,m_1,m_2) + 
                             p_1^\mu p_1^\nu B_{21}(p_1,m_1,m_2)\;.
\end{eqnarray}
\end{subequations}
The integrals $B_{22}$, $B_{21}$ and $B_1$ may be reexpressed in terms of
the $B_0$ function and the one-point integral, $A$\cite{pv79,hhkm94}.  Before 
proceeding it is convenient to introduce some short-hand notation that will be 
useful in the reductions of the $C$-functions.
\begin{subequations}
\begin{eqnarray}
B_n^{(12)} & = & B_n = B_n(p_1,m_1,m_2)\;,\\
B_n^{(13)} & = & B_n(p_1+p_2,m_1,m_3)\;,\\
B_n^{(23)} & = & B_n(p_2,m_2,m_3)\;.
\end{eqnarray}
\end{subequations}
where $B_n = B_{22}$, $B_{21}$, $B_{1}$ or $B_{0}$.

We introduce our $C$-functions by
\begin{equation}
\label{C-def}
\{C_0,\,C^{\mu},\,C^{\mu\nu},\,C^{\mu\nu\rho} \}(p_1,p_2,m_1,m_2,m_3) 
= \int\frac{\overline{d^dk}}{i\pi^2}
\frac{\{1,\,k^\mu,\,k^\mu k^\nu ,\,k^\mu k^\nu k^\rho \}}{N_1N_2N_3}\;,
\end{equation}
and the tensor integrals may be expanded as
\begin{subequations}
\label{C-scalar-def}
\begin{eqnarray}
C^{\mu}(p_1,p_2,m_1,m_2,m_3) & = & p_1^\mu C_{11} + p_2^\mu C_{12}\;,\\
C^{\mu\nu}(p_1,p_2,m_1,m_2,m_3) & = & 
        p_1^\mu p_1^\nu C_{21} + p_2^\mu p_2^\nu C_{22} +
        ( p_1^\mu p_2^\nu + p_2^\mu p_1^\nu ) C_{23} + g^{\mu\nu} C_{24} \;,\\
\nonumber
C^{\mu\nu\rho}(p_1,p_2,m_1,m_2,m_3) & = & 
        p_1^\mu p_1^\nu p_1^\rho C_{31} + p_2^\mu p_2^\nu p_2^\rho C_{32} \\
\nonumber && \mbox{}
     + ( p_1^\mu p_1^\nu p_2^\rho + p_1^\mu p_2^\nu p_1^\rho + 
           p_2^\mu p_1^\nu p_1^\rho ) C_{33}
     + ( p_2^\mu p_2^\nu p_1^\rho + p_2^\mu p_1^\nu p_2^\rho + 
           p_1^\mu p_2^\nu p_2^\rho ) C_{34}\\
&& \mbox{}
  + ( g^{\mu\nu}p_1^\rho + g^{\mu\rho}p_1^\nu + g^{\nu\rho}p_1^\mu ) C_{35}
  + ( g^{\mu\nu}p_2^\rho + g^{\mu\rho}p_2^\nu + g^{\nu\rho}p_2^\mu ) C_{36}\;.
\end{eqnarray}
\end{subequations}
It is convenient to introduce the following $2\times 2$ matrix: 
\begin{equation}
\label{x2}
X = \left( \begin{array}{cc} 2p_1^2 & 2p_1\cdot p_2 \\ 2p_1\cdot p_2 & 2p_2^2
      \end{array} \right)
\end{equation}
Then the reduction of the $C_{1n}$ functions is accomplished by  
\begin{equation}
\label{c1n-reduction}
\left( \begin{array}{c} C_{11} \\ C_{12} \end{array} \right) = 
X^{-1} \left( \begin{array}{l} B_0^{(13)} - B_0^{(23)} + f_1 C_0 \\
    B_0^{(12)} - B_0^{(13)} + f_2 C_0 \end{array} \right)\;.
\end{equation}
For the $C_{2n}$ functions,
\begin{subequations}
\label{c2n-reduction}
\begin{eqnarray}
C_{24} & = & \frac{1}{4} + \frac{1}{4}B_0^{(23)}+\frac{1}{2}m_1^2 C_0 
             - \frac{1}{4} ( f_1 C_{11} + f_2 C_{12} ) \;, \\
\left(\begin{array}{c} C_{21} \\ C_{23} \end{array}\right) & = & X^{-1} 
\left(\begin{array}{l} B_1^{(13)} + B_0^{(23)} + f_1 C_{11} - 2 C_{24} \\
  B_1^{(12)} - B_1^{(13)} + f_2 C_{11} \end{array}\right) \;,\\ 
\left(\begin{array}{c} C_{23} \\ C_{22} \end{array}\right) & = & X^{-1} 
\left(\begin{array}{l} B_1^{(13)} - B_1^{(23)} + f_1 C_{12}  \\
 \makebox[1cm]{}  - B_1^{(13)} + f_2 C_{12} - 2 C_{24} \end{array}\right) \;.
\end{eqnarray}
\end{subequations}
Finally, for the $C_{3n}$ functions,
\begin{subequations}
\label{c3n-reduction}
\begin{eqnarray}
\left(\begin{array}{c} C_{35} \\ C_{36} \end{array}\right) & = & X^{-1} 
\left(\begin{array}{l} 
B_{22}^{(13)} - B_{22}^{(23)} + f_1 C_{24} \\
B_{22}^{(12)} - B_{22}^{(13)} + f_2 C_{24}
\end{array}\right) 
\;, \label{c3i}\\
\left(\begin{array}{c} C_{31} \\ C_{33} \end{array}\right) & = & X^{-1}
\left(\begin{array}{l} 
B_{21}^{(13)} - B_{0}^{(23)} + f_1 C_{21} - 4 C_{35} \\
B_{21}^{(12)} - B_{21}^{(13)} + f_2 C_{21} 
\end{array}\right)  
\;,\\
\left(\begin{array}{c} C_{34} \\ C_{32} \end{array}\right) & = & X^{-1} 
\left(\begin{array}{l} 
B_{21}^{(13)} - B_{21}^{(23)} + f_1 C_{22} \\
\makebox[1cm]{} - B_{21}^{(13)} + f_2 C_{22} - 4 C_{36}  
\end{array}\right) 
\;,\\
\left(\begin{array}{c} C_{33} \\ C_{34} \end{array}\right) & = & X^{-1} 
\left(\begin{array}{l} 
B_{21}^{(13)} + B_{1}^{(23)} + f_1 C_{23} - 2 C_{36} \\
\makebox[1cm]{} - B_{21}^{(13)} + f_2 C_{23} - 2 C_{35}  
\end{array}\right) 
\;.
\end{eqnarray}
\end{subequations}
It is also possible to solve for $C_{35}$ and $C_{36}$ {\em via}
\begin{subequations}
\begin{eqnarray}
\label{c3n-also}
C_{35} & = & -\frac{1}{9} - \frac{1}{6}B_0^{(23)} + \frac{1}{3} m_1^2 C_{11}
             -\frac{1}{6}(f_1 C_{21} + f_2 C_{23}) \;,\\
C_{36} & = & -\frac{1}{18} + \frac{1}{6}B_1^{(23)} + \frac{1}{3} m_1^2 C_{12}
             -\frac{1}{6}(f_1 C_{23} + f_2 C_{22}) \;.
\end{eqnarray}
\end{subequations}

These reductions can sometimes be numerically unstable, especially when 
the matrix of Eqn.~(\ref{x2}) becomes singular.  Using the equations above 
we are able to calculate $C_{33}$, $C_{34}$, $C_{35}$ and $C_{36}$ in two 
different ways.  The redundancy provides a useful check of numerical
reliability.

\newpage


\begin{thebibliography}{99}
%
\bibitem{brs} C.~Becchi, A.~Rouet and B.~Stora, Commun. Math. Phys. {\bf 42} 
              (1975); Ann. Phys. (N.Y.) {\bf 98} (1976) 287.
%
\bibitem{gkn86} G.J.~Gounaris, R.~K\"{o}gerler and H.~Neufeld,
                Phys. Rev. {\bf D34} (1986) 3257.
%
\bibitem{hhis97}  K.~Hagiwara, T.~Hatsukano, S.~Ishihara 
                  and R.~Szalapski, Nucl. Phys. {\bf B496} (1997) 66.
%
\bibitem{helas} H.~Murayama, I.~Watanabe and K.~Hagiwara, HELAS: 
                HELicity Amplitude Subroutines for 
                Feynman diagram evaluations, KEK-91-11, Jan. 1992, 184.   
%
\bibitem{madgraph} T.~Stelzer and W.F.~Long, Comput. Phys. Commun. 81
                   (1994) 357-371.
%
\bibitem{himmz91} K.~Hagiwara, H.~Iwasaki, A.~Miyamoto, H.~Murayama 
                  and D.~Zeppenfeld, Nucl. Phys. {\bf B365} (1991) 544.
%
\bibitem{apls88} C.~Ann, M.E.~Peskin, B.W.~Lynn and S.~Selipsky, Nucl. Phys. 
                 {\bf B309} (1988) 221.
%
\bibitem{et} J.M.~Cornwall, D.N.~Levin and G.Tiktopoulos, Phys. Rev. Lett. 
             {\bf 30} (1973) 1268; Phys. Rev. {\bf D10} (1974) 1145; B.W.~Lee, 
             C.~Quigg and H.B. Thacker, Phys. Rev. {\bf D16} (1977) 1519.
%
\bibitem{et2} M.S. Chanowitz and M.K.~Gaillard, Nucl. Phys. {\bf B261} (1985) 
              379; H.~Veltman, Phys. Rev. {\bf D41} (1990) 2294.  
\bibitem{et3} 
             J.~Bagger and C.~Schmidt, Phys. Rev. {\bf D41} (1990) 264; 
              H.-J. He, Y.-P. Kuang and X. Li, Phys. Rev. Lett. {\bf 69} 
              (1992) 2619; Phys. Rev. {\bf D49} (1994) 4842. 
%
\bibitem{pv79} G.~Passarino and M.~Veltman, Nucl. Phys. {\bf B160} (1979) 151.
%
\bibitem{hisz93} K.~Hagiwara, S.~Ishihara, R.~Szalapski and D.~Zeppenfeld,
                 Phys. Rev. {\bf D48} (1993) 2182.
%
\bibitem{lv80} M.~Lemoine and M.~Veltman, Nucl. Phys. {\bf B164} (1980) 445.
%
\bibitem{fr97} F.~Feruglio and S.~Rigolin, Phys. Lett. {\bf B397} (1997) 
               245-254.
%
\bibitem{hpzh87} K.~Hagiwara, R.D.~Peccei, D.~Zeppenfeld and K.~Hikasa, Nucl. 
                 Phys. {\bf B282} (1987) 253.
%
\bibitem{achksu} S.~Alam, G.C.~Cho, K.~Hagiwara, S.~Kanemura, 
                 R.~Szalapski and Y.~Umeda, {\em in preparation.}
%
\bibitem{hhkm94} K.~Hagiwara, D.~Haidt, C.~S.~Kim and S.~Matsumoto, 
                 Z.~Phys.~{\bf C64} (1994) 559.
%
\bibitem{sk97} S.~Kanemura, KEK-Preprint 97-160 ({\tt hep-ph 9710237}).
%
\bibitem{msbar} O.V.~Tarasov, A.A.~Vladimirov and A.Yu~Zharkov, Phys. Lett. 
                {\bf 93B} (1980) 429; K.G.~Chetyrkin, A.L.~Kataev and 
                F.V.~Tkachov, Nucl. Phys. {\bf B174} (1987) 125.
%
\bibitem{alam94} S.~Alam, Phys. Rev. {\bf D50} (1994) 124; {\em ibid.} 148; 
                 {\em ibid.} 174.
%
\bibitem{gjo91} G.J.~van~Oldenborgh, Comput. Phys. Commun. {\bf 66} (1991) 1.
%
\end{thebibliography}
\end{document}